\documentclass[useAMS,usenatbib,usegraphicx]{mn2e}

\title[Bayesian Re-analysis of the Gliese 581 Exoplanet System]{Bayesian Re-analysis of the Gliese 581 Exoplanet System}
\author[P. C. Gregory]{Philip C. Gregory$^{1}$\thanks{E-mail:
gregory@phas.ubc.ca} \\
$^{1}$Physics and Astronomy Department, University of British Columbia, 6224 Agricultural Rd., Vancouver, BC V6T 1Z1, Canada}

\begin{document}

\date{Submitted to MNRAS December 28, 2010}

\pagerange{\pageref{firstpage}--\pageref{lastpage}} \pubyear{2002}

\maketitle

\label{firstpage}

\begin{abstract}
A re-analysis of Gliese 581 HARPS and HIRES precision radial velocity data was carried out with a Bayesian multi-planet Kepler periodogram (from 1 to 6 planets) based on a fusion Markov chain Monte Carlo algorithm. In all cases the analysis included an unknown parameterized stellar jitter noise term. For the HARPS data set the most probable number of planetary signals detected is 5 with a Bayesian false alarm probability of 0.01. These include the $3.1498\pm0.0005$, $5.3687\pm0.0002$, $12.927_{-0.004}^{+0.006}$, and $66.9\pm0.2$d periods reported previously plus a $399_{-16}^{+14}$d period. The orbital eccentricities are $0.0_{-0.0}^{+0.2}$, $0.00_{-0.00}^{+0.02}$, $0.10_{-0.10}^{+0.06}$, $0.33_{-0.10}^{+0.09}$, and $0.02_{-0.02}^{+0.30}$, respectively. The semi-major axis and $M sin i$ of the 5 planets are ($0.0285\pm0.0006$ au, $1.9\pm0.3$M$_{\earth}$), ($0.0406\pm0.0009$ au, $15.7\pm0.7$M$_{\earth}$), ($0.073\pm0.002$ au, $5.3\pm0.4$M$_{\earth}$), ($0.218\pm0.005$ au, $6.7\pm0.8$M$_{\earth}$), and ($0.7\pm0.2$ au, $6.6_{-2.7}^{+2.0}$M$_{\earth}$), respectively.  
 
The analysis of the HIRES data set yielded a reliable detection of only the strongest 5.37 and 12.9 day periods. The analysis of the combined HIRES/HARPS data again only reliably detected the 5.37 and 12.9d periods. Detection of 4 planetary signals with periods of 3.15, 5.37, 12.9, and 66.9d was only achieved by including an additional unknown but parameterized Gaussian error term added in quadrature to the HIRES quoted errors. The marginal distribution for the sigma of this additional error term has a well defined peak at $1.8\pm0.4$m s$^{-1}$. It is possible that this additional error arises from unidentified systematic effects. We did not find clear evidence for a fifth planetary signal in the combined HIRES/HARPS data set.
\end{abstract}

\begin{keywords}
stars: planetary systems; methods: statistical; methods: data analysis; techniques: radial velocities.
\end{keywords}

\section{Introduction}

A remarkable array of new ground based and space based astronomical tools have finally provided astronomers access to other solar systems with over 500 planets discovered to date, starting from the pioneering work of \citet{Campbell1998}, \citet{Wolszczan1992}, \citet{Mayor1995}, and \citet{Marcy1996}. Recent interest has focused on the Gl 581 planetary system (also designated GJ 581 in the literature). It was already known to harbor three planets, including two super-Earth planets that straddle its habitable zone: Gl 581 b  with a period of 5.37d (Bonfils et al. 2005b), Gl 581 c (period 12.9d) and d (period 82d) \citep{Udry2007}. Armed with additional HARPS data, \citet{Mayor2009} reported the detection of an additional planet – Gl 581e – with a minimum mass of 1.9M$_{\earth}$ and a period of 3.15d. They also corrected previous confusion about the orbital period of Gl 581d (the outermost planet) with a one-year alias at 82 days. The revised period is 66.8d, and positions the semi-major axis inside the habitable zone of the low mass star. \citet{Vogt2010} reported the analysis of the combined HIRES and HARPS data set spanning 11y, claiming the detection of two additional planets Gl 581f \& g. Gl 581f has a period of 433d, a minimum-mass of 7.0M$_{\earth}$, and a semi-major axis of 0.758 au. Gl 581g has a period of 36.6d, a minimum-mass 3.1M$_{\earth}$, and a semi-major axis of 0.146 au. The estimated equilibrium temperature of Gl 581g is 228 K, placing it squarely in the middle of the habitable zone of the star. The \citet{Vogt2010} analysis assumed circular orbits for all 6 planets.

\citealt{Gregory2009} and \citealt{Gregory2010a} presented a Bayesian hybrid or fusion MCMC algorithm that incorporates parallel tempering (PT), simulated annealing and a genetic crossover operation to facilitate the detection of a global minimum in $\chi^2$. This enables the efficient exploration of a large model parameter space starting from a random location. When implemented with a multi-planet Kepler model, it is able to identify any significant periodic signal component in the data that satisfies Kepler's laws and is able to function as a multi-planet Kepler periodogram~\footnote{Following on from the pioneering work on Bayesian periodograms by \citet{Jaynes1987} and \citet{Brett1988}}. In addition, the Bayesian MCMC algorithm provides full marginal parameters distributions. The algorithm includes an innovative adaptive control system that automates the selection of efficient parameter proposal distributions even if the parameters are highly correlated \citealt{Gregory2010b}. A recent application of the algorithm (\citealt{Gregory2010a}) confirmed the existence of a disputed second planet (\citealt{Fischer2002}) in 47 Ursae Majoris (47 UMa) and provided orbital constraints on a possible additional long period planet with a period $\sim 10000$d.

This paper reports the results of a re-analysis of the HARPS \citep{Mayor2009} and HIRES data \citep{Vogt2010} for Gl 581 using the above Bayesian multi-planet Kepler periodogram. Section~\ref{sec:fusion} provides an introduction to our Bayesian approach and describes the adaptive fusion MCMC algorithm. Section~\ref{sec:analysis} gives the model equations and priors. Sections~\ref{sec:HARPS} and \ref{sec:modsel} present the parameter estimation and model selection results for the analysis of the HARPS data alone. Section~\ref{sec:HIRES} is devoted to the analysis of the HIRES data followed by the analysis of the combination of HIRES and HARPS data. The final two sections are devoted to discussion and conclusions.  

\section{The adaptive fusion MCMC}
\label{sec:fusion}

The adaptive fusion~\footnote{In earlier papers the algorithm was referred to as a hybrid MCMC. We subsequently learned that this term already exists in the literature in connection with a Hamiltonian version of MCMC. In this paper we replace the term hybrid by fusion.}  MCMC (FMCMC) is a very general Bayesian nonlinear model fitting program. After specifying the model, $M_i$, the data, $D$, and priors, $I$, Bayes theorem dictates the target joint probability distribution for the model parameters which is given by
\begin{equation}
p(\vec{X}|D,M_{i},I) = C \ p(\vec{X}|M_{i},I) \times p(D|M_{i},\vec{X},I).
\label{eq:orbit25}
\end{equation} 
where $C$ is the normalization constant and $\vec{X}$ represent the set of model parameters. The first term on the RHS of the equation, $p(\vec{X}|M_i,I)$, is the prior probability distribution of $\vec{X}$, prior to the consideration of the current data $D$. The second term, $p(D|\vec{X},M_i,I)$, is called the likelihood and it is the probability that we would have obtained the measured data $D$ for this particular choice of parameter vector $\vec{X}$, model $M_i$, and prior information $I$. At the very least, the prior information, $I$, must specify the class of alternative models (hypotheses) being considered (hypothesis space of interest) and the relationship between the models and the data (how to compute the likelihood). In some simple cases the log of the likelihood is simply proportional to the familiar $\chi^2$ statistic. For further details of the likelihood function for this type of problem see \cite{Gregory2005b}.

To compute the marginals for any subset of the parameters it is necessary to integrate the joint probability distribution over the remaining parameters. For example, the marginal probability density function (PDF) of the orbital period in a one planet radial velocity model fit is given by
\\

\begin{eqnarray}
p(P|D,M_{1},I) & = & \int dK \int dV \int de \int d\chi \int d\omega \int ds\nonumber\\
& & \times \; p(P,K,V,e,\chi,\omega,s|D,M_{1},I) \nonumber\\
& \propto & p(P|M_{1},I) \int dK \cdots \int ds \nonumber\\
& & \times \; p(K,V,e,\chi,\omega,s|M_{1},I) \nonumber\\ 
& & \times \; p(D|M_{1},P,K,V,e,\chi,\omega,s,I),
\label{eq:orbit}
\end{eqnarray} 
where $p(P,K,V,e,\chi,\omega,s|D,M_{1},I)$ is the target joint probability distribution of the radial velocity model parameters ($P,K,V,e,\chi,\omega$) and $s$ is an extra noise parameter which is discussed in Section~\ref{sec:analysis}. $p(P|M_{1},I)$ is the prior for the orbital period parameter, $p(K,V,e,\chi,\omega,s|M_{1},I)$ is the joint prior for the other parameters, and $p(D|M_{1},P,K,V,e,\chi,\omega,s,I)$ is the likelihood. For a five planet model fit we need to integrate over 26 parameters to obtain $p(P|D,M_{1},I)$. Integration is more difficult than maximization, however, the Bayesian solution provides the most accurate information about the parameter errors and correlations without the need for any additional  calculations, i.e., Monte Carlo simulations. Bayesian model selection requires integrating over all the model parameters.

In high dimensions, the principle tool for carrying out the integrals is Markov chain Monte Carlo based on the Metropolis algorithm. The greater efficiency of an MCMC stems from its ability, after an initial burn-in period, to generate  samples in parameter space in direct proportion to the joint target probability distribution. In contrast, straight Monte Carlo integration randomly samples the parameter space and wastes most of its time sampling regions of very low probability. 

MCMC algorithms avoid the requirement for completely independent samples, by constructing a kind of random walk in the model parameter space such that the number of samples in a particular region of this space is proportional to a target posterior density for that region. The random walk is accomplished using a Markov chain, whereby the new sample, $\vec{X}_{t+1}$, depends on previous sample $\vec{X}_t$ according to a time independent entity called the {\it transition kernel}, $p(\vec{X}_{t+1}|\vec{X}_t)$. The remarkable property of $p(\vec{X}_{t+1}|X_t)$ is that after an initial \index{Burn-in period}burn-in period (which is discarded) it generates samples of $\vec{X}$ with a probability density proportional to the desired posterior $p(\vec{X}|D,M_{1},I)$ (e.g., see Chapter 12 of Gregory 2005 for details). 

The transition kernel, $p(\vec{X}_{t+1}|\vec{X}_t)$ is given by 
\begin{equation}
p(\vec{X}_{t+1}|\vec{X}_t) = q(\vec{X}_{t+1}|\vec{X}_t) \alpha(\vec{X}_t,\vec{X}_{t+1}),
\label{eq:transkernal}
\end{equation}
where $\alpha(\vec{X}_t,\vec{X}_{t+1})$ is called the acceptance probability and is given by equation~\ref{eq:acceptP}. This is achieved by proposing a new sample $\vec{X}_{t+1}$ from a {\it proposal distribution}, $q(\vec{X}_{t+1}|\vec{X}_t)$, which is easy to evaluate and is centered on the current sample $\vec{X}_t$. The proposal distribution can have almost any form. A common choice for $q(\vec{X}_{t+1}|\vec{X}_t)$ is a multivariate normal (Gaussian) distribution. With such a proposal distribution, the probability density decreases with distance away from the current sample. The new sample $\vec{X}_{t+1}$ is accepted with a probability $\alpha(\vec{X}_t,\vec{X}_{t+1})$ given by
\begin{equation}
\alpha(\vec{X}_t,\vec{X}_{t+1}) = {\rm min}\left[1,\frac{p(\vec{X}_{t+1}|D,I)}{p(\vec{X}_t|D,I)}\ \frac{q(\vec{X}_t|\vec{X}_{t+1})}{q(\vec{X}_{t+1}|\vec{X}_t)}\right],
\label{eq:acceptP}
\end{equation}
where $q(\vec{X}_t|\vec{X}_{t+1})=q(\vec{X}_{t+1}|\vec{X}_t)$ for a symmetrical proposal distribution.
If the proposal is not accepted the current sample $\vec{X}_t$ is repeated. 
\begin{figure*}
\begin{center}
\includegraphics[width=140mm]{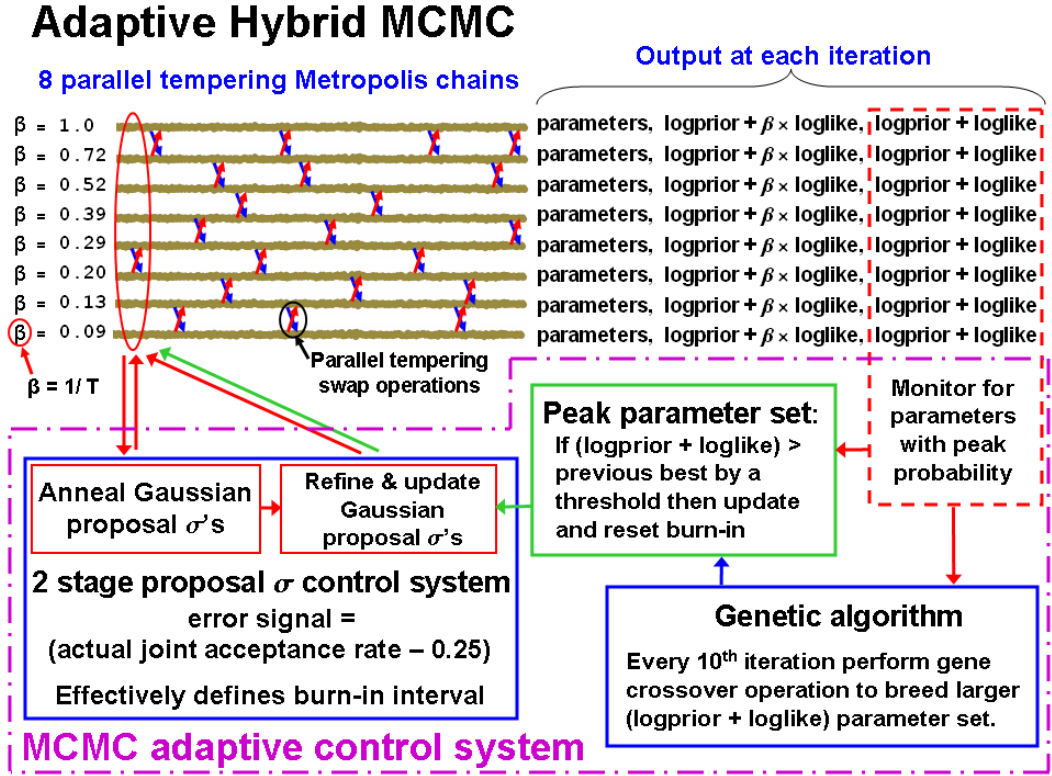}
\caption{The first of two schematics on the operation of the adaptive fusion MCMC (FMCMC) algorithm.}
\end{center}
\label{fig:schem}
\end{figure*}

An important feature that prevents the fusion MCMC from becoming stuck in a local probability maximum is parallel tempering (\citet{Geyer1991} and re-invented by \citet{Hukushima1996} under the name {\it  exchange Monte Carlo}). Multiple MCMC chains are run in parallel. The joint distribution for the parameters ($\vec{X}$) of model $M_i$, for a particular chain, is given by
\begin{equation}
\pi(\vec{X}|D,M_i,I,\beta) \propto p(\vec{X}|M_i,I)\times p(D|\vec{X},M_i,I)^{\beta}.
\label{eq:tempering}
\end{equation}
Each MCMC chain corresponding to a different $\beta$, with the value of $\beta$ ranging from zero to 1. When the exponent $\beta = 1$, the term on the LHS of the equation is the target joint probability distribution for the model parameters, $p(\vec{X}|D,M_i,I)$. For $\beta \ll 1$, the distribution is much flatter.

In equation~\ref{eq:tempering}, an exponent $\beta = 0$ yields a joint density distribution equal to the prior. The reciprocal of $\beta$ is analogous to a temperature, the higher the temperature the broader the distribution. For parameter estimation purposes 8 chains with\\
$\beta=\{0.09, 0.13, 0.20, 0.29, 0.39, 0.52, 0.72, 1.0\}$ were employed. At an interval of 10 iterations, a pair of adjacent chains on the tempering ladder are chosen at random and a proposal made to swap their parameter states. A Monte Carlo acceptance rule determines the probability for the proposed swap to occur (e.g., Gregory 2005a, equation 12.12). This swap allows for an exchange of information across the population of parallel simulations. In low $\beta$ (higher temperature) simulations, radically different configurations can arise, whereas in higher $\beta$ (lower temperature) states, a configuration is given the chance to refine itself. The lower $\beta$ chains can be likened to a series of scouts that explore the parameter terrain on different scales. The final samples are drawn from the $\beta  = 1$ chain, which corresponds to the desired target probability distribution. The choice of $\beta$ values can be checked by computing the swap acceptance rate. When they are too far apart the swap rate drops to very low values. A swap acceptance rate of $\approx 40\%$ works well.

At each iteration, a single joint proposal to jump to a new location in the parameter space is generated from independent Gaussian proposal distributions (centered on the current parameter location), one for each parameter. In general, the $\sigma$'s of these Gaussian proposal distributions are different because the parameters can be very different entities. If the $\sigma$'s are chosen too small, successive samples will be highly correlated and will require many iterations to obtain an equilibrium set of samples. If the $\sigma$'s are too large, then proposed samples will very rarely be accepted. The process of choosing a set of useful proposal $\sigma$'s when dealing with a large number of different parameters can be very time consuming. In parallel tempering MCMC, this problem is compounded because of the need for a separate set of Gaussian proposal $\sigma$'s for each tempering chain. This process is automated by an innovative three stage statistical control system (\citealt{Gregory2007b}, \citealt{Gregory2009}) in which the error signal is proportional to the difference between the current joint parameter acceptance rate and a target acceptance rate, typically 25\% (\citealt{Roberts1997}). A schematic of the first two stages of the adaptive control system (CS) is shown in Fig. 1. A third stage that handles highly correlated parameters is described in Section~\ref{sec:CorPar}.
 
The first stage CS, which involves annealing the set of Gaussian proposal distribution $\sigma$'s, was described in Gregory 2005a. An initial set of proposal $\sigma$'s ($\approx 10\%$ of the prior range for each parameter) are used for each chain. During the major cycles, the joint acceptance rate is measured based on the current proposal $\sigma$'s and compared to a target acceptance rate. During the minor cycles, each proposal $\sigma$ is separately perturbed to determine an approximate gradient in the acceptance rate. The $\sigma$'s are then jointly modified by a small increment in the direction of this gradient. This is done for each of the parallel chains. Proposals to swap parameter values between chains are allowed during major cycles but not within minor cycles. 

\begin{figure}
\includegraphics[width=85mm]{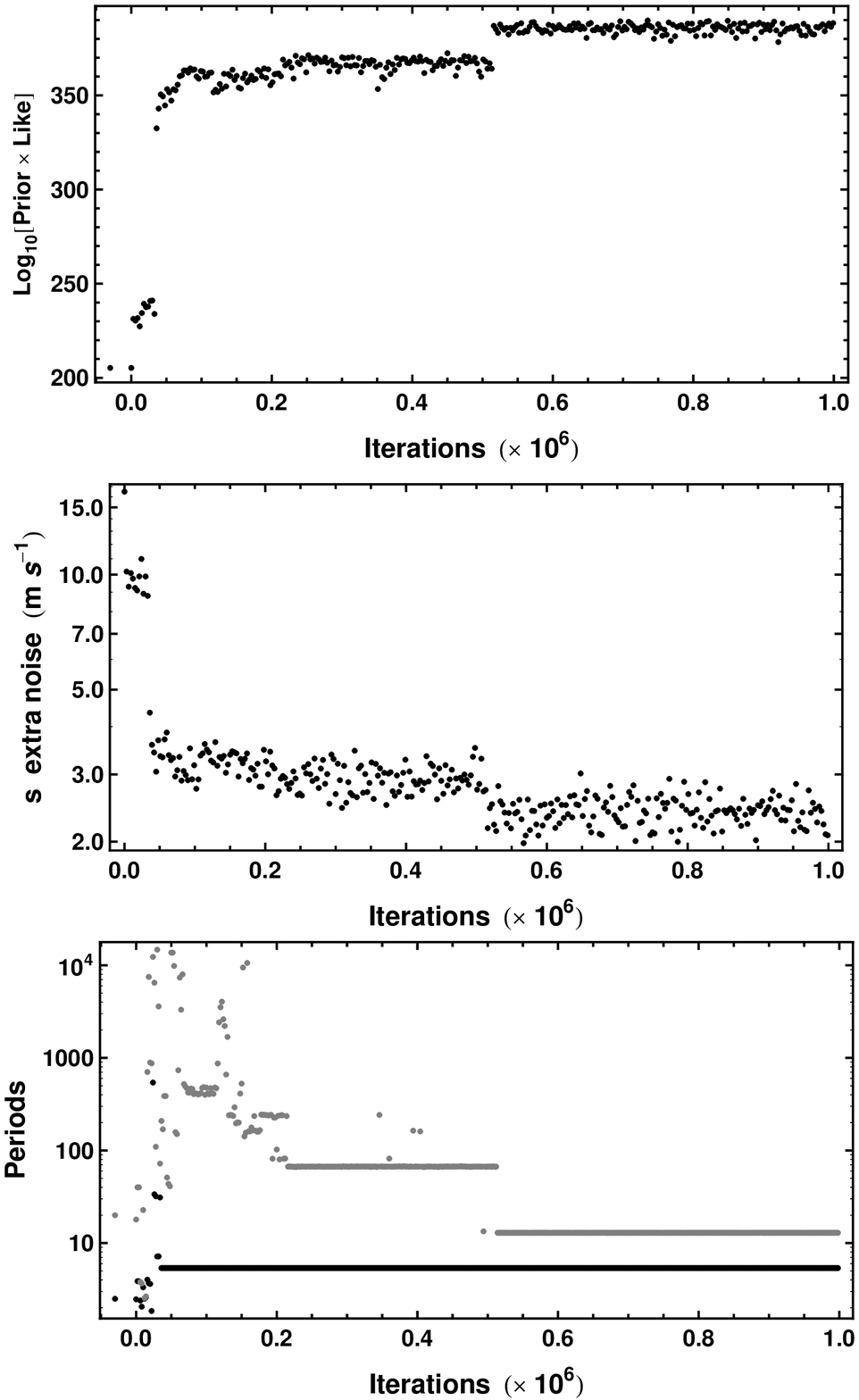}
\caption{The upper panel is a plot of the Log$_{10}$[Prior $\times$ Likelihood] versus MCMC iteration for a blind 2 planet fit to the HARPS data. The middle panel is a similar plot for the extra noise term $s$. Initially $s$ is inflated and then rapidly decays to a much lower level as the better fit parameter values are approached. The lower panel shows the values of the two unknown period parameters versus iteration number. The two starting periods of 2.5 and 20d are shown on the left hand side of the plot at a negative iteration number. }
\label{fig:sAnnealing}
\end{figure}

The annealing of the proposal $\sigma$'s occurs while the FMCMC is homing in on any significant peaks in the target probability distribution. Concurrent with this, another aspect of the annealing operation takes place whenever the Markov chain is started from a location in parameter space that is far from the best fit values. This automatically arises because all the models considered incorporate an extra additive noise term (\citealt{Gregory2005b}) whose probability distribution is Gaussian with zero mean and with an unknown standard deviation $s$. When the $\chi^2$ of the fit is very large, the Bayesian Markov chain automatically inflates $s$ to include anything in the 
data that cannot be accounted for by the model with the current set of 
parameters and the known measurement errors. This results in a smoothing out of the detailed structure in the $\chi^2$ surface and, as pointed out by \citet{Ford2006}, allows the Markov chain to explore the large scale structure in parameter space more quickly. The chain begins to decrease the value of the extra noise as it settles in near the best-fit parameters. An example of this is shown in Fig.~\ref{fig:sAnnealing} for a two planet fit to the HARPS data as discussed in Section~\ref{sec:2HARPS}. The three panels shows the evolution of the Log$_{10}$[Prior $\times$ Likelihood], the $s$ parameter and the two period parameters. In the early stages the extra noise is inflated to around 16m s$^{-1}$ and then rapidly decays to much lower values as it homes in on possible solutions. This is similar to simulated annealing, but does not require choosing a cooling scheme. In this example, the starting parameter values were far from the best and the MCMC algorithm finds several less probable solutions on route to a final best choice. Initially it homes in on the two periods of 5.37 and 66.9d and the CS switches off around iteration 220,000, but around iteration 500,000 it detects a much more probable solution for the period combination of 5.37 and 12.9d. In this example the adaptive control system switched on again briefly following the detection of the much improved solution. 

Although the first stage CS achieves the desired joint acceptance rate, it often happens that a subset of the proposal $\sigma$'s are too small leading to an excessive autocorrelation in the FMCMC iterations for these parameters. Part of the second stage CS corrects for this. 
The goal of the second stage is to achieve a set of proposal $\sigma$'s that equalizes the FMCMC acceptance rates when new parameter values are proposed separately and achieves the desired acceptance rate when they are proposed jointly. Details of the second stage CS were given in \citealt{Gregory2007b}. 

The first stage is run only once at the beginning, but the second stage can be executed repeatedly, whenever a significantly improved parameter solution emerges. Frequently, the algorithm homes in on the most significant peak within the span of the first stage CS  and the second stage improves the choice of proposal $\sigma$'s based on the highest probability parameter set. 
Occasionally, a new higher (by a user specified threshold) target probability parameter set emerges after the first two stages of the CS have completed. The control system has the ability to detect this and automatically re-activate the second stage. In this sense the CS is adaptive. If this happens the iteration corresponding to the end of the control system is reset. The requirement that the transition kernel be time independent means that $q(\vec{X}_{t+1}|\vec{X}_t)$ be time independent, so useful FMCMC simulation data are obtained only after the CS is switched off.

The adaptive capability of the control system can be appreciated from an examination of Fig. 1. The upper left portion of the figure depicts the FMCMC iterations from the 8 parallel chains, each corresponding to a different tempering level $\beta$ as indicated on the extreme left. One of the outputs obtained from each chain at every iteration (shown at the far right) is the $\log$ prior $+ \log$ likelihood. This information is continuously fed to the CS which constantly updates the most probable parameter combination regardless of which chain the parameter set occurred in. This is passed to the `Peak parameter set' block of the CS. Its job is to decide if a significantly more probable parameter set has emerged since the last execution of the second stage CS. If so, the second stage CS is re-run using the new more probable parameter set which is the basic adaptive feature of the existing CS.

The CS also includes a genetic algorithm block which is shown in the bottom right of Fig. 1. The current parameter set can be treated as a set of genes. In the present version, one gene consists of the parameter set that specifies one orbit. On this basis, a three planet model has three genes. At any iteration there exist within the CS the most probable parameter set to date $\vec{X}_{\rm max}$, and the current most probable parameter set of the 8 chains, $\vec{X}_{\rm cur}$. At regular intervals (user specified) each gene from $\vec{X}_{\rm cur}$ is swapped for the corresponding gene in $\vec{X}_{\rm max}$. If either substitution leads to a higher probability it is retained and $\vec{X}_{\rm max}$ updated. The effectiveness of this operation can be tested by comparing the number of times the gene crossover operation gives rise to a new value of $\vec{X}_{\rm max}$ compared to the number of new $\vec{X}_{\rm max}$ arising from the normal parallel tempering FMCMC iterations. The gene crossover operations prove to be very effective, and give rise to new $\vec{X}_{\rm max}$ values $\approx 3$ times more often. Of course, most of these swaps lead to very minor changes in probability but occasionally big jumps are created. 

Gene swaps from $\vec{X}_{\rm cur2}$, the parameters of the second most probable current chain, to $\vec{X}_{\rm max}$ can also be also utilized. This gives rise to new values of $\vec{X}_{\rm max}$ at a rate approximately half that of swaps from $\vec{X}_{\rm cur}$ to $\vec{X}_{\rm max}$. Crossover operations at a random point in the entire parameter set did not prove as effective except in the single planet case where there is only one gene. Further experimentation with this concept is ongoing.    

\subsection{Highly correlated parameters}
\label{sec:CorPar}

\begin{figure*}
\begin{center}
\includegraphics[width=140mm]{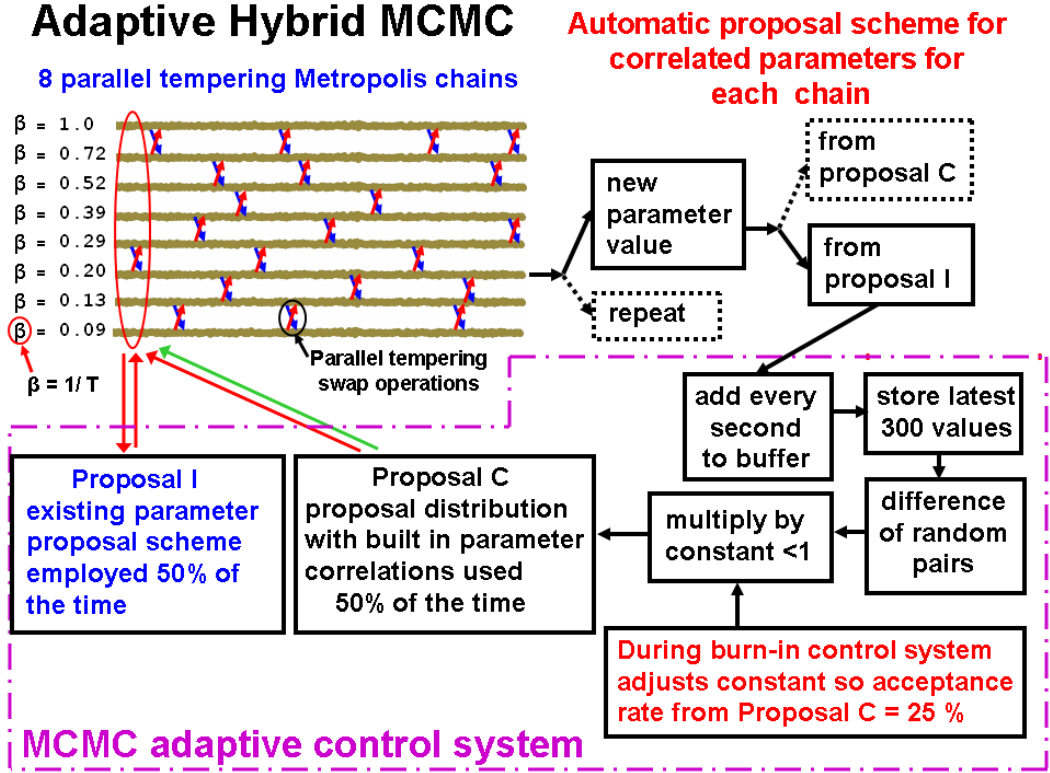}
\caption{Schematic outlining the operation of the third stage of the adaptive fusion MCMC algorithm that handles correlated parameters.}
\end{center}
\label{fig:schem2}
\end{figure*}

The part of the algorithm described above (Figure 1) is most efficient when working with model parameters that are independent or transformed to new independent parameters. New parameter values are jointly proposed based on independent Gaussian proposal distributions (`I' scheme), one for each parameter. 
Initially, only this `I' proposal system is used and it is clear that if there are strong correlations between any parameters the $\sigma$'s of the independent Gaussian proposals will need to be very small for any proposal to be accepted and consequently convergence will be very slow. However, the accepted `I' proposals will generally cluster along the correlation path. In the optional third stage of the control system shown in Figure~\ref{fig:schem2}, every second accepted `I' proposal is appended to a correlated sample buffer. Only the 300 most recent additions to the buffer are retained. A `C' proposal is generated from the difference between a pair of randomly selected samples drawn from the correlated sample buffer, after multiplication by a constant. The value of this constant is computed automatically by another control system module which ensures that the `C' proposal acceptance rate is close to $25\%$. With very little computational overhead, the `C' proposals provide the scale and direction for efficient jumps in a correlated parameter space. 

The final proposal distribution is a random selection of `I' and `C' proposals such that each is employed 50\% of the time. The overhead to generate the `C' proposals is minimal. The combination ensures that the whole parameter space can be reached and that the FMCMC chain is aperiodic. The parallel tempering feature operates as before to avoid becoming trapped in a local probability maximum. 

Because the `C' proposals reflect the parameter correlations, large jumps are possible allowing for much more efficient movement in parameter space than can be achieved by the `I' proposals alone.  Once the first two stages of the control system have been turned off, the third stage continues until a minimum of an additional $300$ accepted `I' proposals have been added to the buffer and the `C' proposal acceptance rate is within the range $\ge 0.22$ and $\le 0.28$. At this point further additions to the buffer are terminated and this sets a lower bound on the burn-in period.

Full details on the operation and testing of the combined `I' and `C' proposal scheme are given in \citealt{Gregory2010b}.

\section{Models and Priors}
\label{sec:analysis}

In this section we describe the model fitting equations and the selection of priors for the model parameters. We have investigated the Gl 581 data using models ranging from 1 to 6 planets. For a one planet model the predicted radial velocity is given by
\begin{equation}
v(t_i) = V + K [\cos\{\theta(t_i+\chi P)+\omega\} + e \cos \omega],
\label{eq:orbit1}
\end{equation}
and involves the 6 unknown parameters
\begin{itemize}
\item[] $V =$ a constant velocity.
\item[] $K =$ velocity semi-amplitude. 
\item[] $P =$ the orbital period.
\item[] $e =$ the orbital eccentricity.
\item[] $\omega =$ the longitude of periastron.
\item[] $\chi =$ the fraction of an orbit, prior to the start of data taking, that periastron occurred at. Thus, $\chi P =$ the number of days prior to $t_i = 0$ that the star was at periastron, for an orbital period of P days. 
\item[] $\theta(t_i+\chi P) =$ the true anomaly, the angle of the star in its orbit relative to periastron at time $t_i$.
\end{itemize}

We utilize this form of the equation because we obtain the dependence of $\theta$ on $t_i$ by solving the conservation of angular momentum equation
\begin{equation}
\frac{d\theta}{dt} - \frac{2 \pi [1+e\cos \theta(t_i+\chi \; P)]^2}{P (1-e^2)^{3/2}} = 0.
\label{eq:orbit2}
\end{equation}
Our algorithm is implemented in {\it Mathematica} and it proves faster for {\it Mathematica} to solve this differential equation than solve the equations relating the true anomaly to the mean anomaly via the eccentric anomaly. {\it Mathematica} generates an accurate interpolating function between $t$ and $\theta$ so the differential equation does not need to be solved separately for each $t_i$. Evaluating the interpolating function for each $t_i$ is very fast compared to solving the differential equation, so the algorithm should be able to handle much larger samples of radial velocity data than those currently available without a significant increase in computational time.
For example, an increase in the data by a factor of 6.5 resulted in only an 18\% increase in execution time.

As described in more detail in \citealt{Gregory2007a}, we employed a re-parameterization of $\chi$ and $\omega$ to improve the MCMC convergence speed motivated by the work of Ford (2006). The two new parameters are $\psi=2\pi\chi+\omega$ and $\phi=2 \pi\chi-\omega$. Parameter $\psi$ is well determined for all eccentricities. Although $\phi$ is not well determined for low eccentricities, it is at least orthogonal to the $\psi$ parameter. We use a uniform prior for $\psi$ in the interval 0 to $4 \pi$ and uniform prior for $\phi$ in the interval $-2 \pi$ to $+2 \pi$. This insures that a prior that is wraparound continuous in $(\chi,\omega)$ maps into a wraparound continuous distribution in $(\psi,\phi)$. To account for the Jacobian of this re-parameterization it is necessary to multiply the Bayesian integrals by a factor of $(4 \pi)^{-nplan}$, where $nplan =$ the number of planets in the model. Also, by utilizing the orthogonal combination $(\psi,\phi)$ it was not necessary to make use of the 'C' proposal scheme outlined in Section~\ref{sec:CorPar} which typically saves about 25\% in execution time.

In a Bayesian analysis we need to specify a suitable prior for each parameter. These are tabulated in Table~\ref{tab:priors}. For the current problem, the prior given in Equation~\ref{eq:tempering} is the product of the individual parameter priors. Detailed arguments for the choice of each prior were given in \citealt{Gregory2007a}.     
\begin{table*}
 \centering
 \begin{minipage}{140mm}
  \caption{Prior parameter probability distributions.}
  \label{tab:priors}
  \begin{tabular}{@{}llll@{}}
  \hline
   Parameter    &    Prior        & Lower bound & Upper bound\\
 \hline
Orbital frequency  & $p(\ln f_1, \ln f_2, \cdots \ln f_n|M_n,I) = \frac{n!}{[\ln (f_H/f_L)]^n}$  & 1/1.1 d & 1/1000 yr  \\
&\ ($n=$number of planets)  & &  \\
& & & \\
Velocity $K_i$  &  Modified Jeffreys~\footnote{Since the prior lower limits for $K$ and $s$ include zero, we used a modified Jeffreys prior of the form
\begin{equation}
p(X|M,I) = \frac{1}{X+X_0}\; \frac{1}{\ln\left(1+\frac{X_{\rm max}}{X_0}\right)}
\label{eq:orbit13}
\end{equation}
For $X \ll X_0$, $p(X|M,I)$ behaves like a uniform prior and for $X \gg X_0$ it behaves like a Jeffreys prior. The $\ln\left(1+\frac{ X_{\rm max}}{X_0}\right)$ term in the denominator ensures that the prior is normalized in the interval 0 to $X_{\rm max}$.} & 0 \ (K$_0 = 1)$ &  $K_{\rm max}\ \left(\frac{P_{\rm min}}{P_i}\right)^{1/3} \frac{1}{\sqrt{1-e_i^2}}$ \\
\ \ \  (m s$^{-1}$) & & & \\
  & \ \ \ \ \ $\frac{(K+K_0)^{-1}}{\ln{\left[1+\frac{K_{\rm max}}{K_0} \ \left(\frac{P_{\rm min}}{P_i}\right)^{1/3} \frac{1}{\sqrt{1-e_i^2}}\right]}}$
 &  & $K_{\rm max}=2129$\\
 & & & \\
V  (m s$^{-1}$) & Uniform & $-K_{\rm max}$ & $K_{\rm max}$  \\
& & & \\
$e_i$ Eccentricity & a) Uniform & 0 & 1 \\
 & b) Ecc. noise bias correction filter& 0 & 0.99 \\
& & & \\
$\chi$ orbit fraction & Uniform & $0$ & 1 \\
& & & \\
$\omega_i$ Longitude of & Uniform & $0$ & $2 \pi$ \\
\ \ \ \ periastron &  &  & \\
& & & \\
$s$ Extra noise   (m s$^{-1}$) & $\frac{(s+s_0)^{-1}}{\ln{\left(1+\frac{s_{\rm max}}{s_{0}}\right)}}$ & 0  \ (s$_0 = 1$)& $K_{\rm max}$  \\
\hline
\end{tabular}
\end{minipage}
\end{table*}

\citealt{Gregory2007a} discussed two different strategies to search the orbital frequency parameter space for a multi-planet model: (i) an upper bound on $f_1 \le f_2 \le \cdots \le f_n$  is utilized to maintain the identity of the frequencies, and (ii) all $f_i$ are allowed to roam over the entire frequency range  and the parameters re-labeled afterwards. Case (ii) was found to be significantly more successful at converging on the highest posterior probability peak in fewer iterations during repeated blind frequency searches. In addition, case (ii) more easily permits the identification of two planets in 1:1 resonant orbits. We adopted approach (ii) in the current analysis. 

All of the models considered in this paper incorporate an extra noise parameter, $s$, that can allow for any additional noise beyond the known measurement uncertainties~\negthinspace\footnote{In the absence of detailed knowledge of the sampling distribution for the extra noise, we pick a Gaussian because for any given finite noise variance it is the distribution with the largest uncertainty as measured by the entropy, i.e., the maximum entropy distribution (\citealt{Jaynes1957}, \citealt{Gregorybook} section 8.7.4.)}. We assume the noise variance is finite and adopt a Gaussian distribution with a variance $s^2$. Thus, the combination of the known errors and extra noise has a Gaussian distribution with variance $= \sigma_i^2 + s^2$, where $\sigma_i$ is the standard deviation of the known noise for i$^{\mbox{\tiny th}}$ data point. For example, suppose that the star actually has two planets, and the model assumes only one is present. In regard to the single planet model, the velocity variations induced by the unknown second planet acts like an additional unknown noise term. Other factors like star spots and chromospheric activity can also contribute to this extra velocity noise term which is often referred to as stellar jitter. Several researchers have attempted to estimate stellar jitter for individual stars based on statistical correlations with observables (e.g., \citealt{Saar1997}, \citealt{Saar1998}, \citealt{Wright2005}). In general, nature is more complicated than our model and known noise terms. Marginalizing $s$ has the desirable effect of treating anything in the data that can't be explained by the model and known measurement errors as noise, leading to conservative estimates of orbital parameters. See Sections 9.2.3 and 9.2.4 of \citet{Gregorybook} for a tutorial demonstration of this point. If there is no extra noise then the posterior probability distribution for $s$ will peak at $s = 0$. The upper limit on $s$ was set equal to $K_{\rm max}$. We employed a modified Jeffrey's prior for $s$ with a knee, $s_0 = 1$m s$^{-1}$. 

We used two different choices of priors for eccentricity, a uniform prior and eccentricity noise bias correction filter that is described in the next section.

\subsection{Eccentricity bias}
\label{sec:eccBias}

In \citealt{Gregory2010a}, the velocities of model fit residuals were randomized in multiple trials and processed using a one planet version of the FMCMC Kepler periodogram. In this situation periodogram probability peaks are purely the result of the effective noise. The orbits corresponding to these noise induced periodogram peaks exhibited a well defined statistical bias towards high eccentricity. They offered the following explanation for this effect. To mimic a circular velocity orbit the noise points need to be correlated over a larger fraction of the orbit than they do to mimic a highly eccentric orbit. For this reason it is more likely that noise will give rise to spurious highly eccentric orbits than low eccentricity orbits. 

\begin{figure}
\includegraphics[width=80mm]{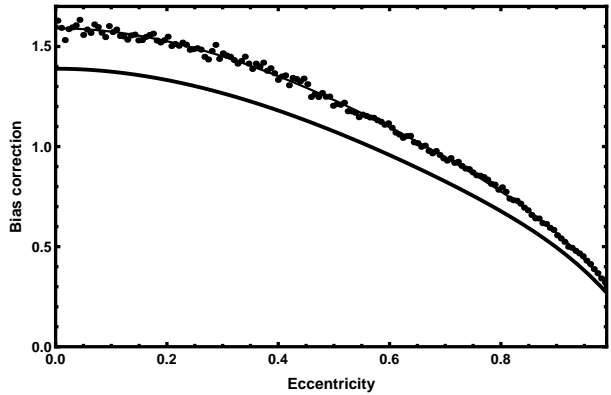}
\caption{The best fit polynomial (thin black curve) to the reciprocal of the mean of the eccentricity bias determined by \citealt{Gregory2010a}. After normalization this yields the eccentricity noise bias correction filter (lower solid black curve).}
\label{fig:EccBias}
\end{figure}

\citealt{Gregory2010a} characterized this eccentricity bias and designed a correction filter that can be used as an alternate prior for eccentricity to enhance the detection of planetary orbits of low or moderate eccentricity. On the basis of our understanding of the mechanism underlying the eccentricity bias, we expect the eccentricity prior filter to be generally applicable to searches for low amplitude orbital signals in precision radial velocity data sets.
The probability density function for this filter is shown by the solid black curve in Figure~\ref{fig:EccBias} and is given by
\begin{equation}
pdf(e) = 1.3889-1.5212 e^2 +0.53944 e^3 -1.6605(e-0.24821)^8.
\label{eq:EccDeBias}
\end{equation}
In a related study, \citet{Shen2009} explored least-$\chi^2$ Keplerian fits to radial velocity data using synthetic data sets. They found that the best fit eccentricities for low signal-to-noise ratio $K/\sigma \le 3$ and moderate number of observations ${\rm N_{obs}}\le 60$, were systematically biased to higher values, leading to a suppression of the number of nearly circular orbits.

In the analysis of the Gl 581 data we used the eccentricity noise bias correction filter as the eccentricity prior on fits of all the models, with occasional runs using a uniform eccentricity prior to test the robustness of our conclusions.

\section{Analysis of the HARPS data}
\label{sec:HARPS}

\begin{figure}
\includegraphics[width=85mm]{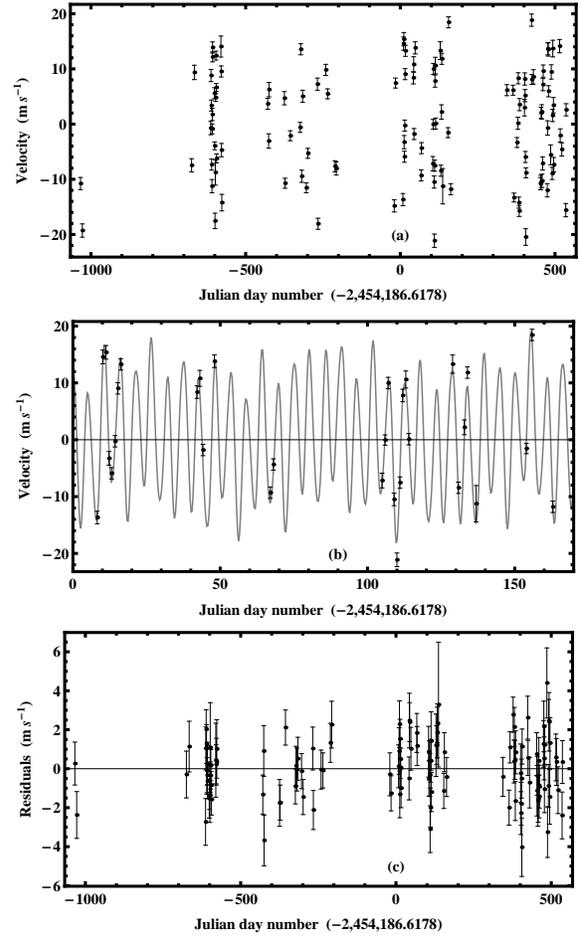}
\caption{Panel (a) shows the HARPS observations of Gl 581. Panel (b) shows a blow-up of the mean 5 planet model fit compared to the data, and panel (c) shows the residuals.}
\label{fig:HARPSdata}
\end{figure}

The HARPS data \citep{Mayor2009} were retrieved electronically~\footnote{ http://cdsweb.u-strasbg.fr/cgi-bin/qcat?J/A+A/507/487}. A mean velocity of -9.2080205km s$^{-1}$ was subtracted and the remainder converted to units of m s$^{-1}$. Panel (a) of 
Figure~\ref{fig:HARPSdata} hows the HARPS observations of Gl 581. Panel (b) shows a blow-up of a portion of the mean 5 planet model fit compared to the data, and panel (c) shows the residuals. The zero reference time is the mean time of the HARPS observations which corresponds to a Julian day number $= 2,454,186.6178$. 

\subsection{Two planet model}
\label{sec:2HARPS}

\begin{figure}
\includegraphics[width=85mm]{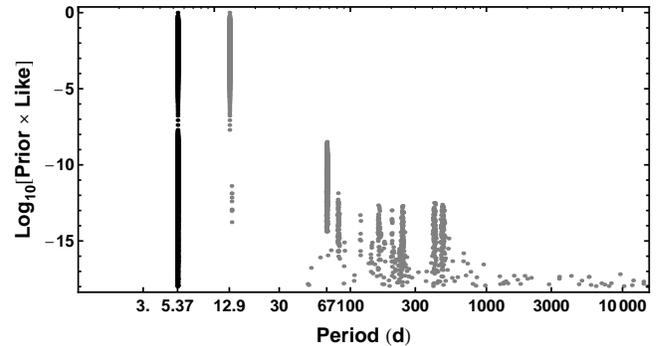}
\caption{A plot of the two period parameter values versus a normalized value of Log$_{10}$[Prior $\times$ Likelihood] for the 2 planet FMCMC Kepler periodogram of the HARPS data.}
\label{fig:2planP}
\end{figure}

The results of our 2 planet Kepler periodogram analysis of this data are shown in Figures~\ref{fig:sAnnealing} and \ref{fig:2planP}. The upper panel of Figure~\ref{fig:sAnnealing} shows a plot of the Log$_{10}$[Prior $\times$ Likelihood] versus FMCMC iteration for a 2 planet fit of the HARPS data. The lower panel shows the values of the two unknown period parameters versus iteration number. The two starting periods of 2.5 and 20d are shown on the left hand side of the plot at a negative iteration number. The larger of the two period parameter finds both the 67 and 12.9d periods but the latter has a much larger Log$_{10}$[Prior $\times$ Likelihood] value. The median value of the extra noise parameter $s = 2.37$m s$^{-1}$.

Figure~\ref{fig:2planP} shows plot of a sample of the FMCMC two period parameters versus a normalized value of Log$_{10}$[Prior $\times$ Likelihood], i.e., a 2 planet periodogram. Only values within 18 decades of the maximum Log$_{10}$[Prior $\times$ Likelihood] are plotted but without regard to whether the values occurred before or after burn-in. The two prominent periods are 5.37 \& 12.9d. The second period parameter exhibited many other peaks but these were all at least 8 decades less probable.  

\subsection{Three planet model}
\label{sec:3HARPS}

\begin{figure}
\includegraphics[width=85mm]{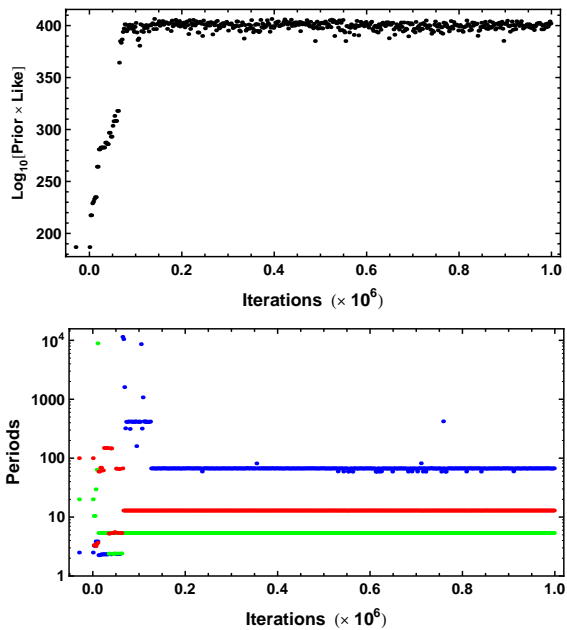}
\caption{The upper panel is a plot of the Log$_{10}$[Prior $\times$ Likelihood] versus iteration for the three planet FMCMC Kepler periodogram of the HARPS data. The lower shows the values of the three unknown period parameters versus iteration number. The three starting periods of 2.5, 20, 100d are shown on the left hand side of the plot at a negative iteration number.}
\label{fig:3planPiter}
\end{figure}

\begin{figure}
\includegraphics[width=85mm]{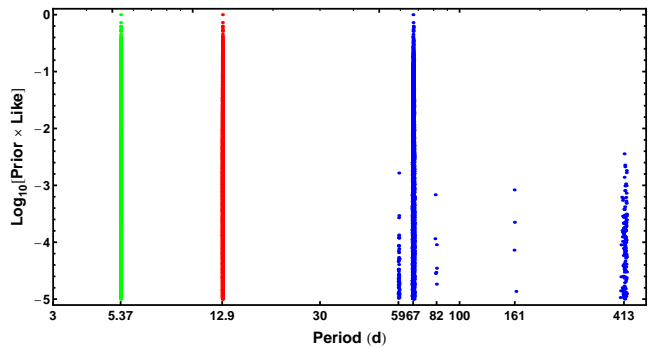}
\caption{A plot of the 3 period parameter values versus a normalized value of Log$_{10}$[Prior $\times$ Likelihood] for the 3 planet FMCMC Kepler periodogram of the HARPS data.}
\label{fig:3planP}
\end{figure}

\begin{figure}
\includegraphics[width=85mm]{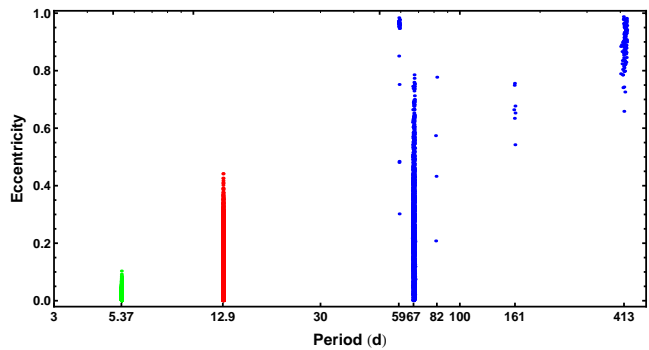}
\caption{A plot of eccentricity versus period for the 3 planet FMCMC Kepler periodogram of the HARPS data.}
\label{fig:3planEccP}
\end{figure}

\begin{figure}
\includegraphics[width=85mm]{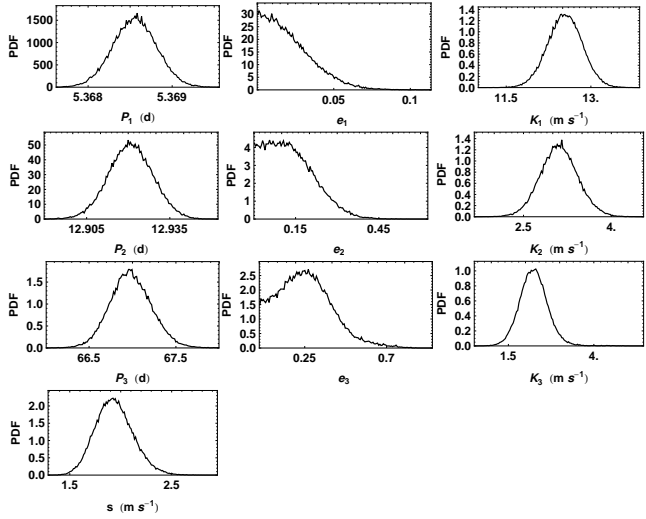}
\caption{A plot of a subset of the FMCMC parameter marginal distributions for a 3 planet fit of the HARPS data.}
\label{fig:3planMarg}
\end{figure}

The results of our 3 planet Kepler periodogram analysis are shown in Figures~\ref{fig:3planPiter}, \ref{fig:3planP}, \ref{fig:3planEccP}, \& \ref{fig:3planMarg}. The upper panel of Figure~\ref{fig:3planPiter} shows a plot of the Log$_{10}$[Prior $\times$ Likelihood] versus FMCMC iteration for a 3 planet fit of the HARPS data. The lower panel shows the FMCMC values of the three unknown period parameters versus iteration number. The three starting periods of 2.5, 20, 100d are shown on the left hand side of the plot at a negative iteration number. The burn-in period for this run was $0.13 \times 10^6$ iterations. 

Figure~\ref{fig:3planP} shows plot of a sample of the FMCMC three period parameters versus a normalized version of Log$_{10}$[Prior $\times$ Likelihood], i.e., a 3 planet periodogram. Only values within 5 decades of the maximum Log$_{10}$[Prior $\times$ Likelihood] are plotted but without regard to whether the values occurred before or after burn-in. Three prominent periods were clearly detected: 5.37, 12.9, 66.9d. The third period parameter exhibited four other peaks but these were all more than 2 decades less probable. The most probable of these has a period $\sim 413$d. The spectral peak at 82d coincides with a one year alias $(1/67-1/365 \sim 1/82)$ of the dominant 67d period. For more on RV aliases see \citet{Dawson2010}. The 59d peak is close but not coincident with the other one year alias $(1/67+1/365 \sim 1/57)$.

Figure~\ref{fig:3planEccP} shows a plot of eccentricity versus period for a sample of the FMCMC parameter samples for the 3 planet model. There is clearly a large uncertainty in the eccentricity of the 67d period which extends down to low eccentricities.
The 413d period peak exhibits very large eccentricity values. \citet{Gregory2010a} showed that it is more likely that noise will give rise to spurious highly eccentric orbits than low eccentricity orbits. To mimic a circular velocity orbit the noise points need to be correlated over a larger fraction of the orbit than they do to mimic a highly eccentric orbit. Even though we are using the noise induced eccentricity prior proposed in \citet{Gregory2010a} we still observe a preponderance of high eccentricity orbital solutions in the low $K$ value regime.

Figure~\ref{fig:3planMarg} shows a plot of a subset of the FMCMC parameter marginal distributions for the 3 planet fit of the HARPS data after filtering out the post burn-in FMCMC iterations that correspond to the 3 dominant period peaks at 5.37, 12.9, and 66.9d. The bottom panel shows the marginal for the unknown standard deviation, $s$, of the additive Gaussian extra noise term which has a median value of 1.93 m s$^{-1}$. The Bayesian analysis automatically inflates $s$ to account for anything in the data that the model and quoted measurement errors cannot account for including stellar jitter.   

\begin{figure}
\includegraphics[width=85mm]{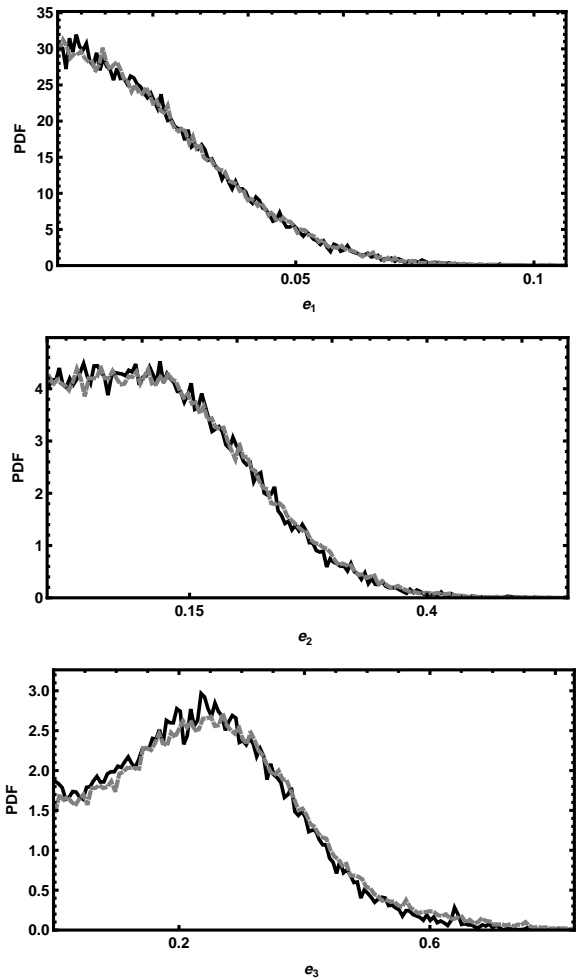}
\caption{The eccentricity marginals for (a) the noise induced eccentricity prior (solid black curve) and (b) the uniform prior (grey dashed curve).}
\label{fig:3planEccNoiseFlatMarg}
\end{figure}

The three planet model was also run using a flat uniform eccentricity to compare with the results obtained with the noise induced eccentricity prior. Figure~\ref{fig:3planEccNoiseFlatMarg} shows a comparison of the eccentricity marginals for the noise induced prior (solid black curve) and the uniform prior (grey dashed curve).    

\subsection{Four planet model}
\label{sec:4HARPS}

\begin{figure}
\includegraphics[width=85mm]{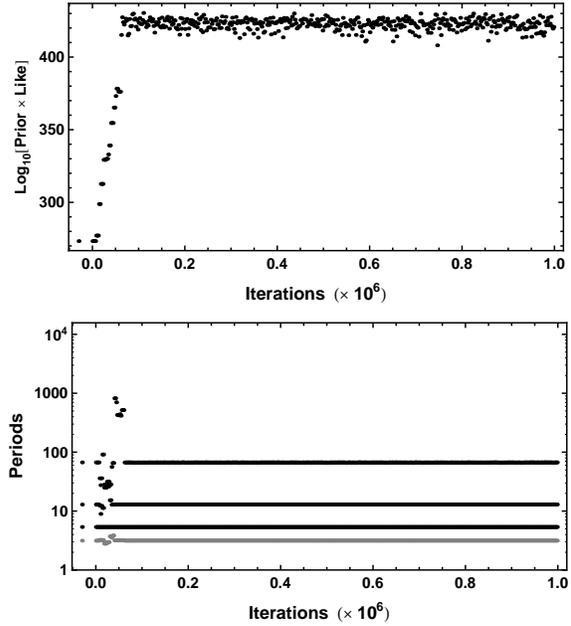}
\caption{The upper panel is a plot of the Log$_{10}$[Prior $\times$ Likelihood] versus iteration for the four planet FMCMC Kepler periodogram of the HARPS data. The lower shows the values of the four unknown period parameters versus iteration number.}
\label{fig:4planPiter}
\end{figure}

\begin{figure}
\includegraphics[width=85mm]{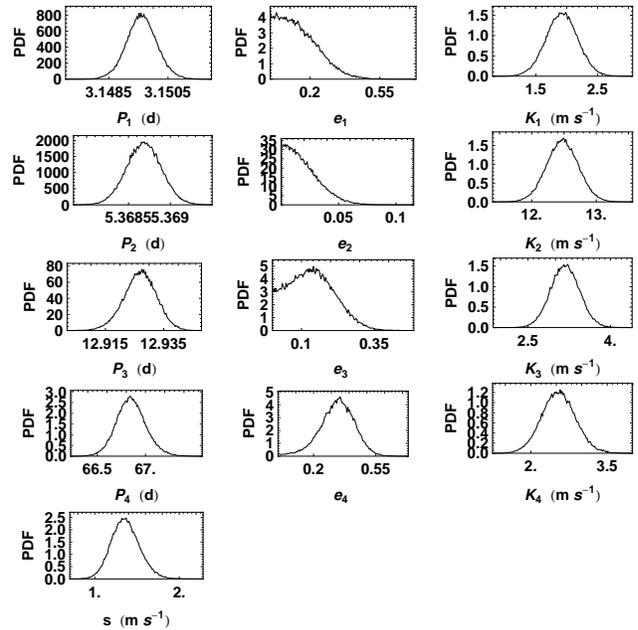}
\caption{A plot of a subset of the FMCMC parameter marginal distributions for a 4 planet fit of the HARPS data.}
\label{fig:4planMarg}
\end{figure}

A one planet fit to the residuals of the three planet fit above yielded a dominant Keplerian orbit with a period of 3.15d. The results of our 4 planet Kepler periodogram analysis are shown in Figures~\ref{fig:4planPiter} and \ref{fig:4planMarg}. 
All 4 period parameters were free to roam within a search range extending from 1.1d to $10 \times$ the data duration. Another run that extended the period search range down to 0.5d yielded the same 4 periods.
The median value of extra noise parameter $s= 1.36$m s$^{-1}$.    

\subsection{Five planet model}
\label{sec:5HARPS}

\begin{figure}
\includegraphics[width=85mm]{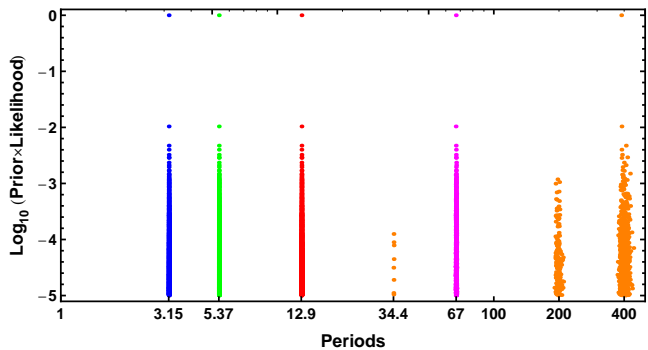}
\caption{A plot of the 5 period parameter values versus a normalized value of Log$_{10}$[Prior $\times$ Likelihood] for the 5 planet FMCMC Kepler periodogram of the HARPS data. The fifth period parameter points are shown in orange.}
\label{fig:5planP}
\end{figure}

\begin{figure}
\includegraphics[width=85mm]{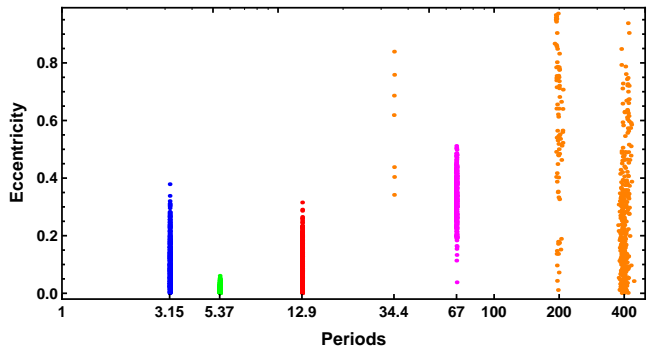}
\caption{A plot of eccentricity versus period for the 5 planet FMCMC Kepler periodogram of the HARPS data. The fifth period parameter points are shown in grey.}
\label{fig:5planEccP}
\end{figure}

\begin{figure}
\includegraphics[width=85mm]{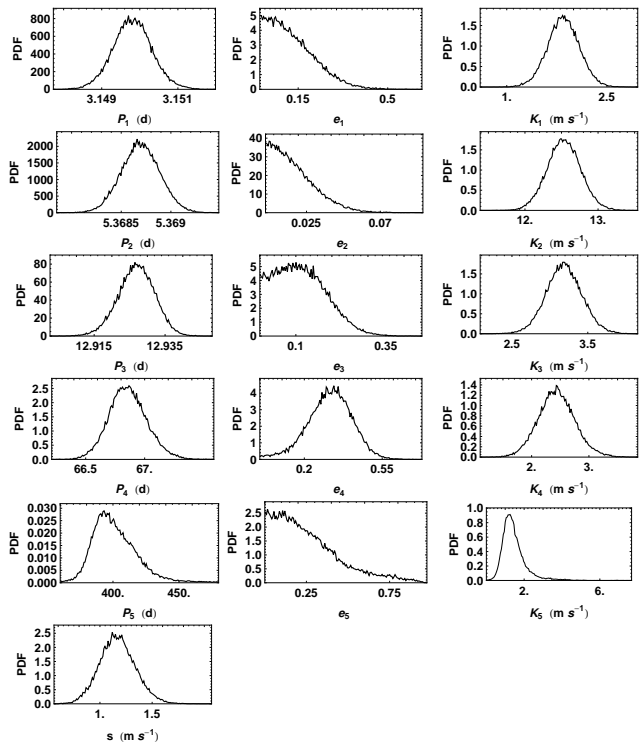}
\caption{A plot of a subset of the FMCMC parameter marginal distributions for a 5 planet fit of the HARPS data.}
\label{fig:5planMarg}
\end{figure}

The results of a 5 planet Kepler periodogram analysis of the HARPS data are shown in Figures~\ref{fig:5planP}, \ref{fig:5planEccP}, and \ref{fig:5planMarg}. The starting period for the fifth period was set $= 300$d and the most probable period found to be $\sim400$d. The best set of parameters from the 4 planet fit were used as start parameters. 
The fifth period parameter shows 3 peaks but the 400d period is almost 1000 times stronger than the other two. We filtered the 5 planet MCMC results to only include fifth period values from the 400d peak region and used the \citet{Gel} statistic to test for convergence. In parallel tempering MCMC, new widely separated parameter values are passed up the line to the $\beta = 1$ simulation and are occasionally accepted. Roughly every 100 iterations the $\beta = 1$ simulation accepts a swap proposal from its neighboring simulation. 
The final $\beta = 1$ simulation is thus an average of a very large number of independent $\beta = 1$ simulations. We divide the $\beta = 1$ iterations into 10 equal time intervals and inter-compared the 10 different essentially independent average distributions for each parameter using a Gelman-Rubin test. For the five planet model results the Gelman-Rubin statistic was $\le 1.01$.
 
Figure~\ref{fig:5planMarg} shows a plot of a subset of the FMCMC parameter marginal distributions for the 5 planet fit of the HARPS data after filtering out the post burn-in FMCMC iterations that correspond to the 5 dominant period peaks at 3.15, 5.37, 12.9, 66.9, and 400d. The median value of the extra noise parameter $s= 1.16$m s$^{-1}$.    


\subsection{Six planet model}
\label{sec:6HARPS}

\begin{figure}
\includegraphics[width=85mm]{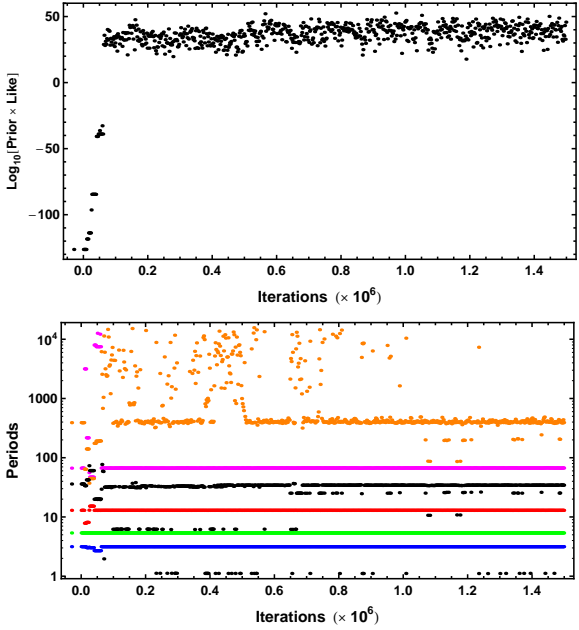}
\caption{The upper panel is a plot of the Log$_{10}$[Prior $\times$ Likelihood] versus iteration for the FMCMC six planet fit of the HARPS data. The lower shows the values of the six unknown period parameters versus iteration number. The six starting periods are shown on the left hand side of the plot at a negative iteration number.}
\label{fig:6planPiter}
\end{figure}

\begin{figure}
\includegraphics[width=85mm]{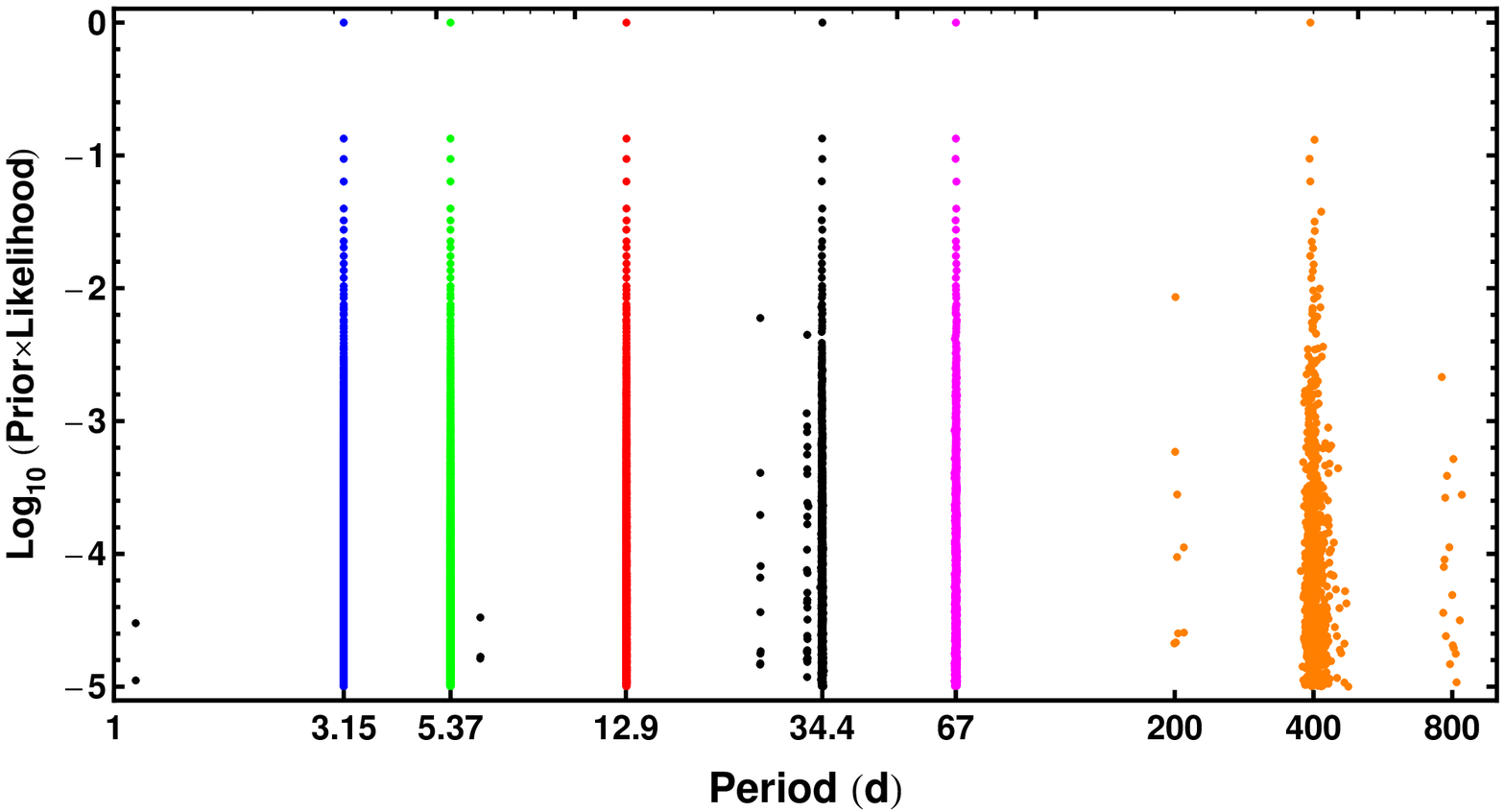}
\caption{A plot of the 6 period parameter values versus a normalized value of Log$_{10}$[Prior $\times$ Likelihood] for the 6 planet FMCMC Kepler periodogram of the HARPS data. The fifth period parameter points are shown in orange and the sixth in black.}
\label{fig:6planP}
\end{figure}

\begin{figure}
\includegraphics[width=85mm]{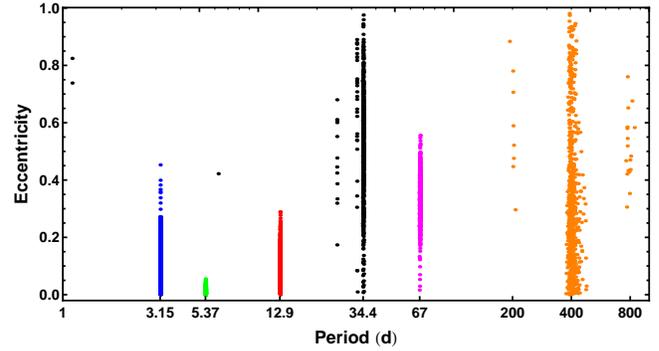}
\caption{A plot of eccentricity versus period for the 6 planet FMCMC Kepler periodogram of the HARPS data.}
\label{fig:6planEccP}
\end{figure}

\begin{figure}
\includegraphics[width=85mm]{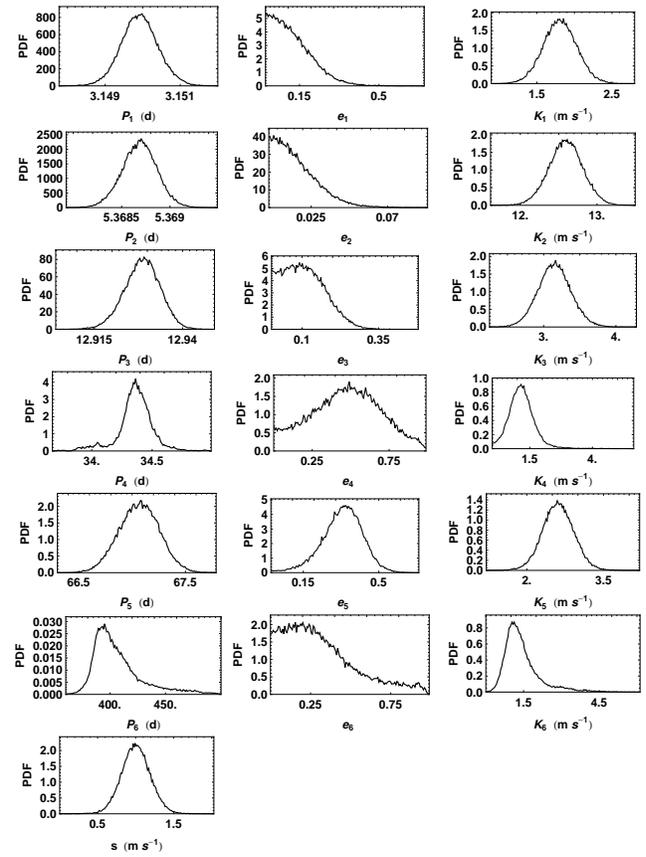}
\caption{A plot of a subset of the FMCMC parameter marginal distributions for a 6 planet fit to the HARPS data.}
\label{fig:6planMarg}
\end{figure}

We also carried out a 6 planet Kepler periodogram analysis of the HARPS data and the results are shown in Figures~\ref{fig:6planPiter}, \ref{fig:6planP}, \ref{fig:6planEccP}, and \ref{fig:6planMarg}. The best set of parameters from the 5 planet fit were used as start parameters and the starting period for sixth period was set $= 36$d. The most probable sixth period found was $34.4$d. A 34.4d period also appeared as a secondary peak in the 5 planet fit and is evident in Figure~\ref{fig:5planP}. Figure~\ref{fig:6planEccP} shows a plot of eccentricity versus period for the 5 planet FMCMC Kepler periodogram.     

Figure~\ref{fig:6planMarg} shows a plot of a subset of the FMCMC parameter marginal distributions for the 6 planet fit of the HARPS data after filtering out the post burn-in FMCMC iterations that correspond to the 6 dominant period peaks at 3.15, 5.37, 12.9, 34.4, 66.9, and 400d. The median value of extra noise parameter $s= 1.00$m s$^{-1}$. There is considerable agreement with the 4 and 5 planet marginals shown earlier which leads us to conclude that the 66.9d orbit is significantly eccentric with an $e \approx 0.34$. There is an indication that $e \approx 0.12$ for the 12.9d orbit. If the 34.4d orbit is real it would also appear to have a significant eccentricity of $0.49_{-0.17}^{+0.22}$.

\begin{figure}
\includegraphics[width=85mm]{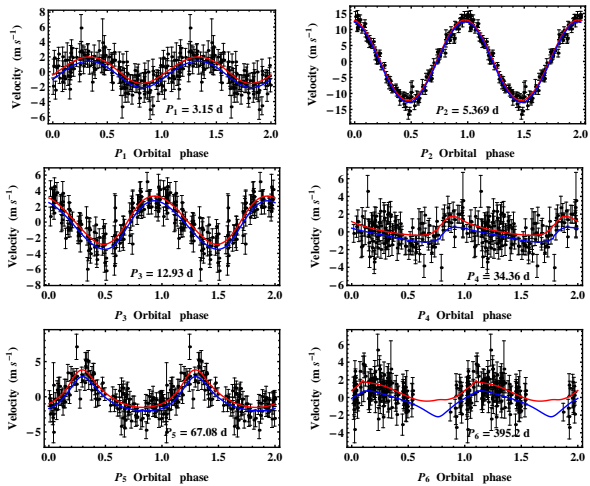}
\caption{Phase plots for the 6 planet model fit to the HARPS data. The top left panel shows the data and model fit versus 3.15 day orbital phase after removing the effects of the five other orbital periods. The red and blue curves are the mean FMCMC model fit $\pm 1$ standard deviation. The other five panels correspond to phase plot for the other five periods.}
\label{fig:6planPhasePls}
\end{figure}

Phase plots for the 6 planet model are shown in Figure~\ref{fig:6planPhasePls}. The top left panel shows the data and model fit versus 3.15d orbital phase after removing the effects of the five other orbital periods. To construct this phase plot we first filter out the post burn-in FMCMC iterations that correspond to the 6 dominant period peaks at 3.15, 5.37, 12.9, 34.4, 66.9, and 400d. The FMCMC output for each of these iterations is a vector of the 6 planet orbital parameter set plus $V$. To compute the 3.15d phase plot data we subtract the mean velocity curve for the other 5 planets plus $V$ from the measured set of velocities. This is done by taking a sample of typically 200 FMCMC iterations and for each iteration we compute the predicted velocity points for that realization of the 5 planet plus $V$ parameter set. We then construct the average of these model prediction data sets and subtract that from the data points. These residuals for the set of observation times are converted to residuals versus phase using the mode of the marginal distribution for the 3.15d period parameter. An orbital phase model velocity fit is then computed at 100 phase points for each realization of the 3.15d planet parameter set obtained in the same sample of 200 iterations as above. At each of these 100 phase points we construct the mean model velocity fit and mean $\pm 1$ standard deviation. The red and blue solid curves in Figure~\ref{fig:6planPhasePls} are the mean FMCMC model fit $\pm 1$ standard deviation. Thus, 68.3\% of the FMCMC model fits fall between these two curves.

The other panels correspond to phase plot for the other five periods. In each panel the quoted period is the mode of the marginal distribution. It is clear that for the 3.15, 5.37, 12.9, 66.9d periods the separation of the fit curves are small compared to the amplitude. For the 395d period phase plot, the wide range of possible orbits that can fit between the red and blue curves is reflected by the broad extent of the marginal distributions of the parameters.

\section{Model Selection for HARPS Analysis}
\label{sec:modsel}

One of the great strengths of Bayesian analysis is the built-in Occam's razor. More complicated models contain larger numbers of parameters and thus incur a larger Occam penalty, which is automatically incorporated in a Bayesian model selection analysis in a quantitative fashion (see for example, \citealt{Gregorybook}, p. 45). The analysis yields the relative probability of each of the models explored.

To compare the posterior probability of the i$^{\rm th}$ planet model to the four planet model we need to evaluate the odds ratio,
$O_{i4} =p(M_{i} | D,I)/p(M_{4} | D,I)$, the ratio of the posterior probability
of model $M_{i}$ to model $M_{4}$.  Application of Bayes 
theorem leads to,
\begin{equation}
O_{i4} = {p(M_{i} | I) \over p(M_{4} | I)}\;
      {p(D | M_{i},I) \over p(D | M_{4},I)}
       \equiv {p(M_{i} | I) \over p(M_{4} | I)}\; B_{i4}
\label{eq:orbit22}
\end{equation}
where the first factor is the prior odds ratio, and the second factor
is called the {\it Bayes factor}, $B_{i4}$. The Bayes factor is the ratio of
the marginal (global) likelihoods of the models. The marginal likelihood for model $M_i$ is given by
\begin{equation}
p(D|M_i,I)= \int d\vec{X} p(\vec{X}|M_i,I)\times p(D|\vec{X},M_i,I).
\label{eq:marglike}
\end{equation}
Thus Bayesian model selection relies on the ratio of marginal likelihoods, not maximum likelihoods. The marginal likelihood is the weighted average of the conditional likelihood, weighted by the prior probability distribution of the model parameters and $s$. This procedure is referred to as marginalization.  

The marginal likelihood can be expressed as the product of the maximum likelihood and the Occam penalty (see \citealt{Gregorybook}, page 48). The Bayes factor will favor the more complicated model only if the maximum likelihood ratio is large enough to overcome this penalty. In the simple case of a single parameter with a uniform prior of width $\Delta X$, and a centrally peaked likelihood function with characteristic width $\delta X$, the Occam factor is $\approx \delta X/\Delta X$. If the data is useful then generally $\delta X \ll \Delta X$. For a 
model with $m$ parameters, each parameter will contribute a term to the overall Occam penalty. The Occam penalty depends not only on the number of parameters but also on the prior range of each parameter (prior to the current data set, $D$), as symbolized in this simplified discussion by $\Delta X$. If two models have some parameters in common then the prior ranges for these parameters will cancel in the calculation of the Bayes factor. To make good use of Bayesian model selection, we need to fully specify priors that are independent of the current data $D$. The sensitivity of the marginal likelihood to the prior range depends on the shape of the prior and is much greater for a uniform prior than a Jeffreys prior (e.g., see \citealt{Gregorybook}, page 61). In most instances we are not particularly interested in the Occam factor itself, but only in the relative probabilities of the competing models as expressed by the Bayes factors. Because the Occam factor arises automatically in the marginalization procedure, its effect will be present in any model selection calculation. Note: no Occam factors arise in parameter estimation problems. Parameter estimation can be viewed as model selection where the competing models have the same complexity so the Occam penalties are identical and cancel out. 
   
The MCMC algorithm produces samples which are in proportion to the posterior probability distribution which is fine for parameter
estimation but one needs the proportionality constant for estimating the model marginal likelihood. 
\citet{Clyde2006} reviewed the state of techniques for model selection from a statistics perspective and \citet{FordGregory2006} have evaluated the performance of a variety of marginal likelihood estimators in the exoplanet context. 

\begin{table*}
  \caption{Marginal likelihood estimates, Bayes factors relative to model 4, and false alarm probabilities. The last two columns list the MAP estimate of the extra noise parameter, $s$, and the RMS residual.}
  \label{tab:modelSel}
  \begin{tabular}{@{}llllllll@{}}
  \hline
   Model & Periods &  Marginal & Bayes factor & False Alarm & $s$ &RMS residual \\
         & (d) &  Likelihood & \ \ nominal & Probability &  (m s$^{-1})$ & (m s$^{-1}$)\\
\hline
$M_{0}$ & & $ 6.10\times 10^{-197}$ & $2.0\times 10^{-59}$ & &  & 9.8\\
& & & & & &\\
$M_{1}$ & $(5.37)$& $(4.221\pm 0.003)\times 10^{-155}$ & $1.4\times 10^{-17}$ & $1.4\times 10^{-42}$ & 3.5 & 3.6\\
& & & & & &\\
$M_{2}$  & $(5.37,12.9)$& $(1.94\pm 0.01)\times 10^{-145}$ & $6.5\times 10^{-8}$ & $2.2\times 10^{-10}$ & 2.4 & 2.6\\
& & & & & &\\
$M_{3}$ & $(5.37,12.9,66.9)$ & $(3.0_{-0.5}^{+0.7}) \times 10^{-142}$ & $10^{-4}$ & $6.5\times 10^{-4}$ & 1.9 & 2.2\\
& & & & & &\\
$M_{4}$  &$(3.15,5.37,12.9,66.9)$& $(3.0_{-0.6}^{+1.1}) \times 10^{-138}$ & $1.0$ & $10^{-4}$ & 1.4 & 1.7\\
& & & & & &\\
$M_{5}$  &$(3.15,5.37,12.9,66.9,399)$& $(3.0_{\times 0.65}^{\times 2.1}) \times 10^{-136}$ & $10^{2}$ & $0.01$ & 1.2 & 1.5\\& & & & & &\\
$M_{6}$  &$(3.15,5.37,12.9,34.4,66.9,399)$& $(6.7^{\times 2.4}_{\times 1/3}) \times 10^{-141}$ & $2.2\times 10^{-3}$ & $0.999978$ & 1.0 & 1.4\\
\hline
\end{tabular}
\end{table*}
Estimating the marginal likelihood is a very big challenge for models with large numbers of parameters, e.g., our six planet model has 32 parameters. In this work we employ the nested restricted Monte Carlo (NRMC) method described in \citealt{Gregory2010a} to estimate the marginal likelihoods. Monte Carlo (MC) integration can be very inefficient in exploring the whole prior parameter range because it randomly samples the whole volume. The fraction of the prior volume of parameter space containing significant probability rapidly declines as the number of dimensions increase. For example, if the fractional volume with significant probability is 0.1 in one dimension then in 32 dimensions the fraction might be of order $10^{-32}$. In restricted MC integration (RMC) this is much less of a problem because the volume of parameter space sampled is greatly restricted to a region delineated by the outer borders of the marginal distributions of the parameters for the particular model. 

In nested RMC (NRMC) integration, multiple boundaries are constructed based on credible regions ranging from 30\% to $\ge 99\%$, as needed. We are then able to compute the contribution to the total integral from each nested interval and sum these contributions. For example, for the interval between the 30\% and 60\% credible regions, we generate random parameter samples within the 60\% region and reject any sample that falls within the 30\% region. Using the remaining samples we can compute the contribution to the NRMC integral from that interval. 

\begin{figure*}
\includegraphics[width=160mm]{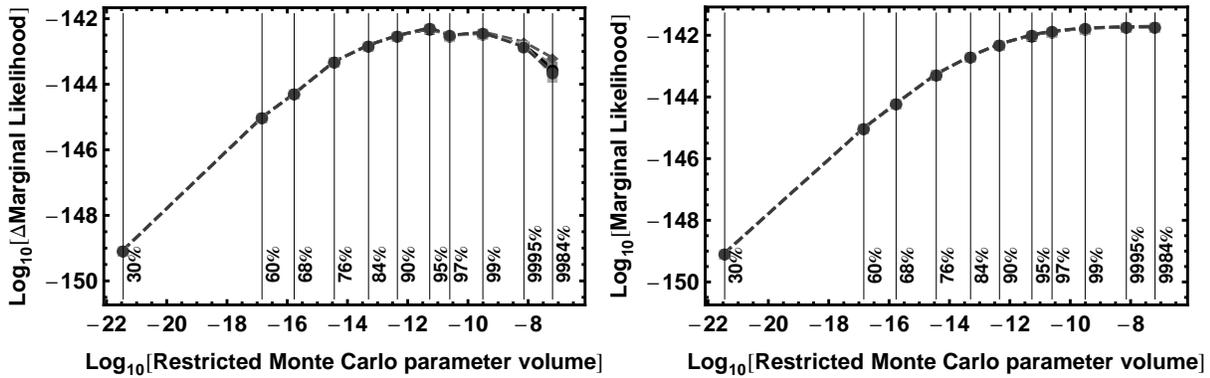}
\caption{Left panel shows the contribution of the individual nested intervals to the NRMC marginal likelihood for the 3 planet model. The right panel shows the integral of these contributions versus the parameter volume of the credible region.}
\label{fig:3planNestedRMC}
\end{figure*}

The left panel of Figure~\ref{fig:3planNestedRMC} shows the contributions from the individual intervals for 5 repeats of the NRMC evaluation for the 3 planet model. The right panel shows the summation of the individual contributions versus the volume of the credible region. The credible region listed as 9995\% is defined as follows. Let $X_{U99}$ and $X_{L99}$ correspond to the upper and lower  boundaries of the 99\% credible region, respectively, for any of the parameters. Similarly, $X_{U95}$ and $X_{L95}$ are the upper and lower boundaries of the 95\% credible region for the parameter. Then $X_{U9995} = X_{U99}+(X_{U99}-X_{U95})$ and $X_{L9995} = X_{L99}+(X_{L99}-X_{L95})$. Similarly, $X_{U9984} = X_{U99}+(X_{U99}-X_{U84})$. 

The NRMC method is expected to underestimate the marginal likelihood in higher dimensions and this underestimate is expected to become worse the larger the number of model parameters, i.e. increasing number of planets (\citealt{Gregory2007c}). When we conclude, as we do, that the NRMC computed odds in favor of the five planet model compared to the four planet model is $\sim 10^{2}$ we mean that the true odds is $\ge 10^{2}$. Thus the NRMC method is conservative. One indication of the break down of the NRMC method is the increased spread in the results for repeated evaluations. 
 
We can readily convert the Bayes factors to a Bayesian False Alarm Probability (FAP) which we define in equation~\ref{eq:FAP1}. For example, in the context of claiming the detection of $m$ planets the FAP$_m$ is the probability that there are actually fewer than $m$ planets, i.e., $m-1$ or less.
\begin{equation}
{\rm FAP}_m = \sum_{i=0}^{m-1}({\rm prob. of \ i \ planets}) 
\label{eq:FAP1}
\end{equation}

If we assume {\it a priori} (absence of the data) that all models under consideration are equally likely, then probability of each model is related to the Bayes factors by
\begin{equation}
p(M_i\mid D,I) = {{B_{i4}} \over {\sum_{j=0}^{N} B_{j4}}}
\label{eq:FAP2}
\end{equation}
where $N$ is the maximum number of planets in the hypothesis space under consideration, and of course
$B_{44} = 1$. For the purpose of computing FAP$_m$ we set $N = m$. Given the Bayes factors in Table~\ref{tab:modelSel} and substituting into equation~\ref{eq:FAP1} gives
\begin{equation}
{\rm FAP}_5 = {{(B_{04} + B_{14} + B_{24}+ B_{34}+ B_{44})}\over {\sum_{j=0}^{5} B_{j4}}}\approx 10^{-2}
\label{eq:FAP3}
\end{equation}
For the 5 planet model we obtain a low FAP $\approx 10^{-2}$. The Bayesian false alarm probabilities for 1, 2, 3, 4, 5, and 6 planet models are given in the fourth column of Table~\ref{tab:modelSel}.

Table~\ref{tab:modelSel} gives the NMRC Marginal likelihood estimates, Bayes factors and false alarm probabilities for 0, 1, 2, 3, 4, 5 and 6 planet models which are designated $M_0, \cdots, M_6$. The last two columns list the maximum {\it a posteriori} (MAP) estimate of the extra noise parameter, $s$, and the RMS residual. For each model the NRMC calculation was repeated 5 times and the quoted errors give the spread in the results, not the standard deviation. The Bayes factors that appear in the third column are all calculated relative to model 4. 

A summary of the 5 planet model parameters and their uncertainties are given in Table~\ref{tab:parerrorsM5}. The quoted value is the median of the marginal probability distribution for the parameter in question (except eccentricity which uses the mode) and the error bars identify the boundaries of the 68.3\% credible region~\footnote{In practice, the probability density for any parameter is represented by a finite list of values $p_i$ representing the probability in discrete intervals $\delta X$. A simple way to compute the 68.3\% credible region, in the case of a marginal with a single peak, is to sort the $p_i$ values in descending order and then sum the values until they approximate 68.3\%, keeping track of the upper and lower boundaries of this region as the summation proceeds.}. The value immediately below in parenthesis is the MAP estimate, the value at the maximum of the joint posterior probability distribution. It is not uncommon for the MAP estimate to fall close to the borders of the credible region. 
In one case, the eccentricity of the fifth planet, the MAP estimate falls well outside the 68.3\% credible region which is one reason why we prefer to quote median or mode values as well. 
The semi-major axis and $M \sin i$ values are derived from the model parameters assuming a stellar mass of $0.31\pm0.02$ M$_{\sun}$ \citep{Delfosse2000}. The quoted errors on the semi-major axis and $M \sin i$ include the uncertainty in the stellar mass. 

\begin{table*}
 \centering
 \begin{minipage}{140mm}
  \caption{Five planet model parameter estimates from HARPS analysis.}
  \label{tab:parerrorsM5}
  \begin{tabular}{@{}llllll@{}}
  \hline
   Parameter  & planet 1 & planet 2 & planet 3 & planet 4 & planet 5 \\
\hline
$P$  (d) & $3.1498_{-.0005}^{+.0005}$ & $5.3687_{-.0002}^{+.0002}$& $12.927_{-.004}^{+.006}$& $66.85_{-0.16}^{+0.15}$& $399_{-16}^{+14}$  \\
& (3.14977)& (5.36866)& (12.9316) & (66.747) & (387.6) \\
& & & & & \\
$K$ (m s$^{-1}$) & $1.85_{-0.22}^{+0.24}$ & $12.53_{-0.22}^{+0.23}$ & $3.18_{-0.24}^{+0.22}$ & $2.43_{-0.31}^{+0.31}$ & $1.3_{-0.5}^{+0.4}$  \\
& (1.93) & (12.39) & (3.40)& (2.75)& (1.62) \\
& & & & & \\
$e$ & $0.11_{-0.11}^{+0.06}$ & $0.015_{-.014}^{+.007}$  & $0.11_{-0.11}^{+0.05}$ & $0.32_{-0.09}^{+0.10}$ & $0.21_{-0.21}^{+0.11}$  \\
& (0.197) & (0.022) & (0.155) & (0.38) & (0.79) \\
& & & & & \\
$\omega$  (deg) & $133_{-75}^{+81}$ & $40_{-82}^{+98}$ &  $234_{-43}^{+43}$ &  $334_{-23}^{+25}$ &  $281_{-100}^{+77}$ \\
& (140) & (-1) & (234) & (326) & (310) \\
& & & & & \\
$a$  (au) & $0.0285_{-.0006}^{+.0006}$ & $0.0406_{-.0009}^{+.0009}$ &  $0.0730_{-.0016}^{+.0016}$ &  $0.218_{-.005}^{+.005}$ &  $0.72_{-0.24}^{+0.24}$\\
& (0.0285) & (0.406) & (0.730) & (0.218) & (0.71)  \\
& & & & & \\
$M \sin i$  ($M_E$) & $1.91_{-0.25}^{+0.26}$ & $15.7_{-0.7}^{+0.7}$ &  $5.29_{-0.43}^{+0.43}$ &  $6.7_{-0.8}^{+0.8}$ &  $6.6_{-2.7}^{+2.0}$ \\
& (1.984) & (15.50) & (5.63) & (7.38) & (5.14) \\
& & & & & \\
Periastron & $4182.6_{-0.7}^{+0.6}$ & $4182_{-1.2}^{+1.4}$ &  $4168.9_{-1.4}^{+1.6}$ &  $4137_{-3.8}^{+3.5}$ &  $3803_{-114}^{+82}$  \\
\ passage &  (4184) & (4186)& (4184) & (4134) & (3828) \\
\ (JD - 2,450,000) & & & & & \\
\hline
\end{tabular}
\end{minipage}
\end{table*}

Although the NRMC estimate of the Bayes factor for the 6 planet model is much lower than for either the 4 or 5 planet models we can still infer the orbital parameters of the most probable additional planetary signal in the 6 planet fit. The period $= 34.4\pm0.1$d, the eccentricity $= 0.49_{-0.17}^{+0.22}$, and the semi-major axis and $M sin i$ are ($0.140\pm0.003$ au, $2.3_{-0.7}^{+0.8}$M$_{\earth}$).   

\section{Analysis of the HIRES data and combination of HIRES and HARPS}
\label{sec:HIRES}

\begin{figure}
\includegraphics[width=85mm]{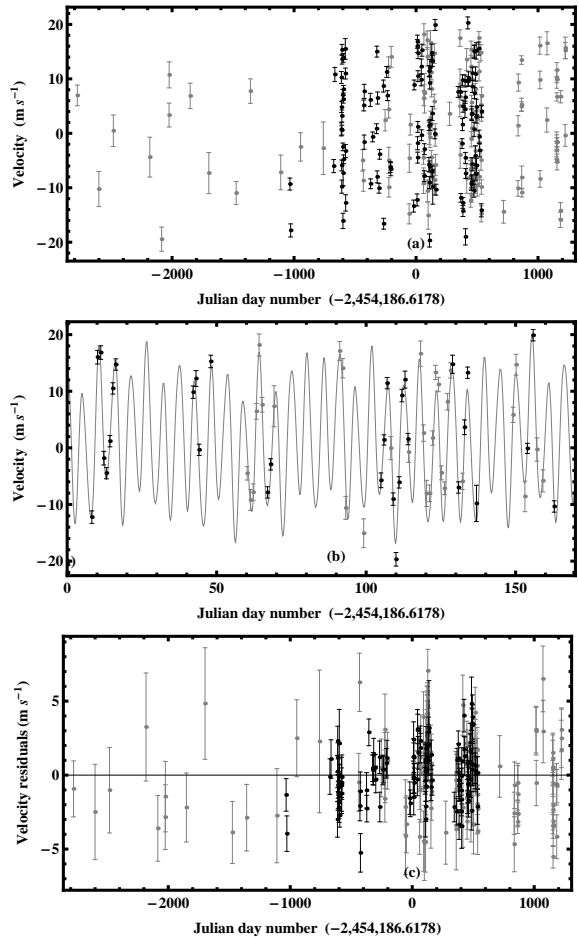}
\caption{Panel (a) shows the combined HIRES (grey points) and HARPS (black points) data set for Gl 581. Panel (b) shows a blow-up of the mean 4 planet model fit compared to the data, and panel (c) shows the residuals.}
\label{fig:HIRESHARPSdata}
\end{figure}

\begin{figure}
\includegraphics[width=85mm]{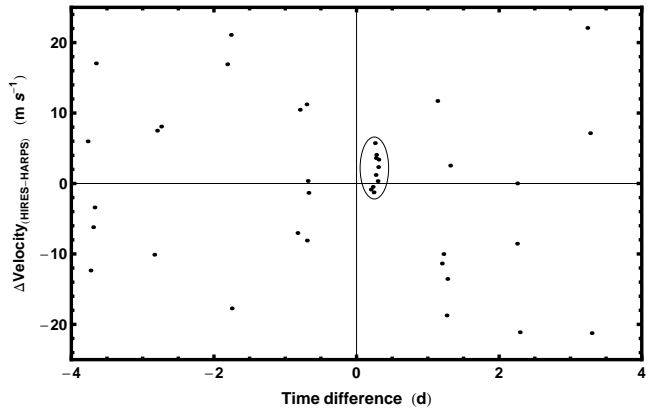}
\caption{A comparison of the velocity differences (HIRES-HARPS) for the nearest pairs of samples versus the sample time difference.}
\label{fig:velDiff}
\end{figure}

In this section we present results on fits to the HIRES data alone and the combination of HIRES and HARPS data.
Panel (a) of Figure~\ref{fig:HIRESHARPSdata} shows the combined HIRES (grey points) and HARPS (black points) data for Gl 581. Panel (b) shows a blow-up of a portion of the mean 4 planet model fit compared to the data, and panel (c) shows the residuals. The same reference time was used for the combined data set as for the HARPS only data. The HIRES data \citep{Vogt2010} consisted of 122 velocity measurements spanning a range of 11 years and with quoted errors ranging from 0.53 to 4.82 m s$^{-1}$. Figure~\ref{fig:velDiff} shows a comparison of the velocity differences (HIRES-HARPS) for the nearest pairs of samples versus the sample time difference. A cluster of 10 points with sample time differences between 0.2 and 0.3s is indicated by the ellipse. Since these time differences are small compared with the shortest known orbital period of 3.15 days, they provide an indication of the agreement between the two sets of measurements. The mean velocity difference for these 10 samples is 1.8 m s$^{-1}$. 
 
The standard deviation of the velocity differences for these 10 pairs is 2.37 m s$^{-1}$. For comparison the mean value of the quoted errors for each pair added in quadrature was 1.94 m s$^{-1}$. 

\subsection{Two planet fit to HIRES data}
\label{sec:2HIRES}

\begin{figure}
\includegraphics[width=85mm]{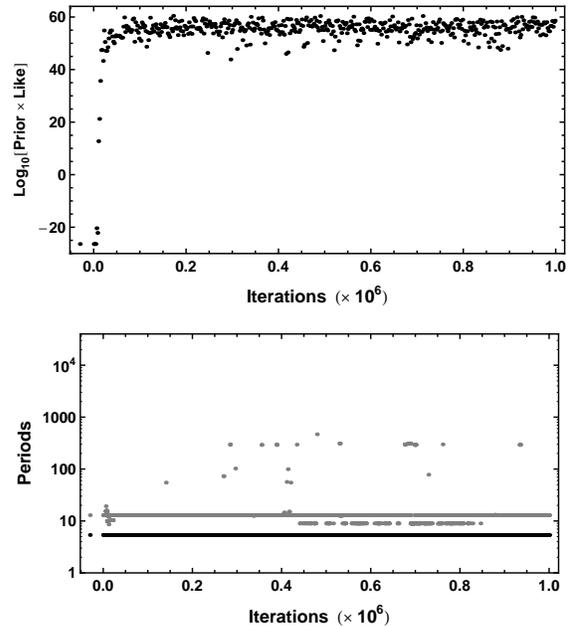}
\caption{The upper panel is a plot of the Log$_{10}$[Prior $\times$ Likelihood] versus iteration for a sample of the FMCMC two planet fit of the HIRES data. The lower shows values of the two unknown period parameters versus iteration number. The two starting periods of 5.37 and 12.9d are shown on the left hand side of the plot at a negative iteration number.}
\label{fig:2planPiterHIRES}
\end{figure}

\begin{figure}
\includegraphics[width=85mm]{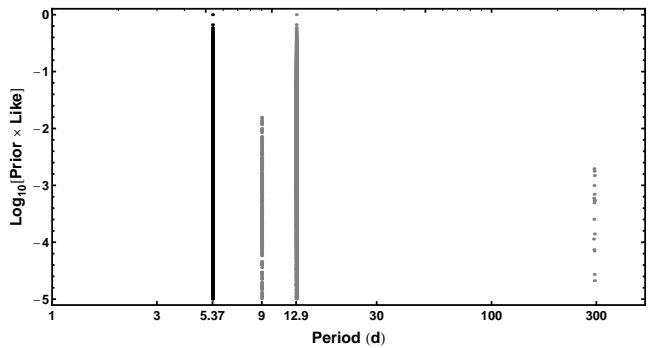}
\caption{A plot of the 2 period parameter values versus a normalized value of Log$_{10}$[Prior $\times$ Likelihood], the 2 planet Kepler periodogram of the HIRES data.}
\label{fig:2planPHIRES}
\end{figure}

\begin{figure}
\includegraphics[width=85mm]{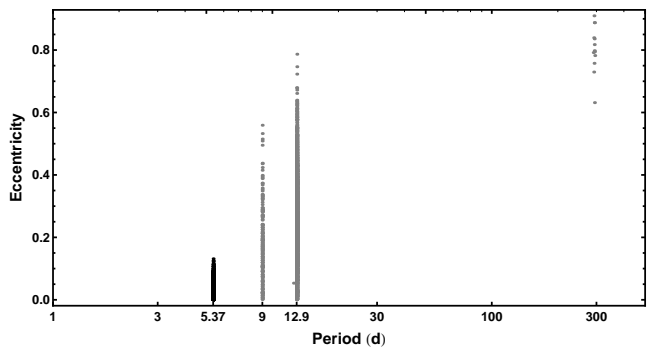}
\caption{A plot of eccentricity versus period for the 2 planet FMCMC fit of the HIRES data.}
\label{fig:2planEccPHIRES}
\end{figure}

The results of our 2 planet Kepler periodogram analysis are shown in Figures~\ref{fig:2planPiterHIRES}, \ref{fig:2planPHIRES}, and \ref{fig:2planEccPHIRES}. The upper panel of Figure~\ref{fig:2planPiterHIRES} shows a plot of the Log$_{10}$[Prior $\times$ Likelihood] versus FMCMC iteration for a 2 planet fit of the HIRES data. The lower panel shows the values of the two unknown period parameters versus iteration number. The two starting periods of 5.37 and 12.9d are shown on the left hand side of the plot at a negative iteration number. The median value of the extra noise parameter $s = 2.69$m s$^{-1}$ compared to $2.37$m s$^{-1}$ for the HARPS 2 planet fit.
Figure~\ref{fig:2planPHIRES} shows the two planet Kepler periodogram. Only values within 5 decades of the maximum Log$_{10}$[Prior $\times$ Likelihood] are plotted but without regard to whether the values occurred before or after burn-in. Two prominent periods were clearly detected: 5.37 and 12.9d. The second period parameter exhibited two other peaks but these were significantly less probable. The most probable of these has a period $\sim 9$d. 

Figure~\ref{fig:2planEccPHIRES} shows a plot of eccentricity versus period for a sample of the FMCMC parameter samples for the 2 planet model. The dominant 5.37 and 12.9d peaks and the weaker 9d peak allow for low eccentricity orbits. The peak around 300d has a high value of eccentricity typical of noise.
  
\subsection{Three planet fit to HIRES data}
\label{sec:3HIRES}

Two three planet runs were carried out on the HIRES only data starting with the best periods (5.37,12.9, 66.9d) found from the HARPS analysis but only the 5.37d period (largest amplitude) was successfully detected. The best of these two runs detected three dominant periods of 5.37, 8.99 and$\sim 300$d. A much weaker peak was found at a period of 12.9d. The HIRES fit extra noise parameter was $s = 2.2$ m s$^{-1}$ compared to the HARPS 3 planet fit where $s = 1.7$ m s$^{-1}$.  
\begin{figure}
\includegraphics[width=85mm]{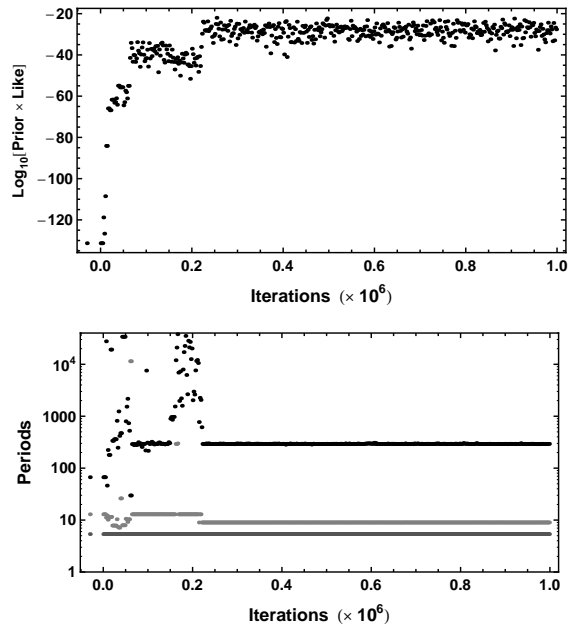}
\caption{The upper panel is a plot of the Log$_{10}$[Prior $\times$ Likelihood] versus iteration for a sample of the FMCMC three planet fit of the HIRES data. The lower shows the values of the three unknown period parameters versus iteration number. The starting periods of 5.37, 12.9 and 66.9d are shown on the left hand side of the plot at a negative iteration number.}
\label{fig:3planPiterHIRES}
\end{figure} 
Figure~\ref{fig:3planPerHIRES} shows a plot of the three planet Kepler periodogram. Only values within 5 decades of the maximum Log$_{10}$[Prior $\times$ Likelihood] are plotted but without regard to whether the values occurred before or after burn-in. Three prominent periods were clearly detected: 5.37, 9, and 300d. The second period parameter (shown in grey) exhibited a second much less probable peak at 12.9d and the third period parameter (black) exhibited many weak peaks. 
\begin{figure}
\includegraphics[width=85mm]{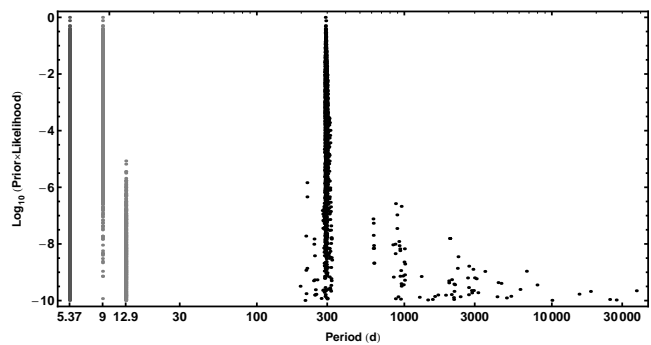}
\caption{A plot of the 3 period parameter values versus a normalized value of Log$_{10}$[Prior $\times$ Likelihood], the 3 planet Kepler periodogram of the HIRES data.}
\label{fig:3planPerHIRES}
\end{figure} 

\begin{figure}
\includegraphics[width=85mm]{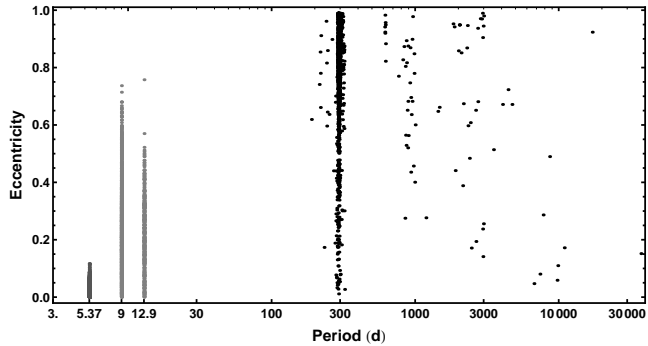}
\caption{A plot of eccentricity versus period for the 3 planet FMCMC fit of the HIRES data.}
\label{fig:3planEccPHIRES}
\end{figure}

Figure~\ref{fig:3planEccPHIRES} shows a plot of eccentricity versus period for a sample of the FMCMC parameter samples for the 3 planet model. The 5.37 and 9d peaks exhibit low eccentricity orbits. The peak around 300d has a high value of eccentricity.

\begin{figure}
\includegraphics[width=85mm]{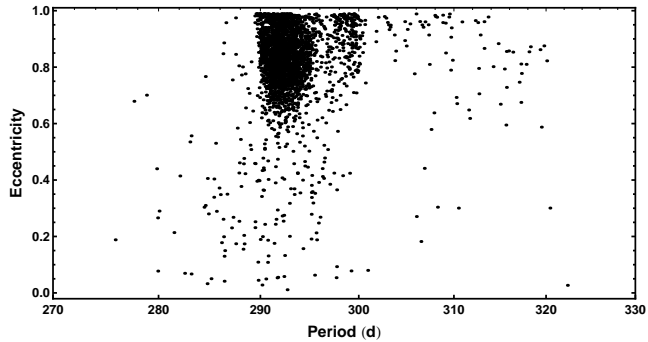}
\caption{A blow-up of the above eccentricity versus period plot in the vicinity of the 300d peak.}
\label{fig:3planEccPblowupHIRES}
\end{figure}

Figure~\ref{fig:3planEccPblowupHIRES} shows a blow-up of the above eccentricity versus period plot in the vicinity of the 300d peak which is dominated by two high eccentricity features typical of noise. We conclude that there is no clear evidence for a third period in the HIRES data alone and suspect that the presence of the strong high eccentricity 300d complex may contribute to the dominance of the 9d period over the 12.9 day period found in the 2 planet fit. 

For both the 2 planet and 3 planet fits the extra noise parameter is larger for the HIRES data than the HARPS data. One possibility is that the quoted HIRES errors have been systematically under-estimated which we can model by an extra Gaussian noise term added in quadrature to the quoted HIRES errors with a $\sigma = ds_{\rm HIRES}$. We can obtain a crude estimate of $ds_{\rm HIRES}$ from the HIRES and HARPS $s$ parameter values for the 2 planet case where both analyses yielded the same 2 periods. The result is $ds_{\rm HIRES}=\sqrt{2.69^2-2.37^2}= 1.3$ m s$^{-1}$. This suggest that in the analysis of the combined data set we should include an extra Gaussian noise term added in quadrature to the quoted HIRES errors with the $\sigma$, labeled $ds_{\rm HIRES}$, as an additional unknown parameter.

\subsection{Two planet fit to the combined HIRES/HARPS data}
\label{sec:2HIRESHARPS}

Based on the above results we decided to use the following noise model for the $j^{\rm{th}}$ data point for the combined HIRES/HARPS two planet analysis.
\begin{equation}
\sigma_{j_{\rm{HARPS}}} = \sqrt{\sigma_{j_{\rm{quoted}}}^2 + s^2}
\label{eq:sigHARPS}
\end{equation} 
\begin{equation}
\sigma_{j_{\rm{HIRES}}} = \sqrt{\sigma_{j_{\rm{quoted}}}^2 + ds_{\rm{HIRES}}^2+ s^2}
\label{eq:sigHIRES}
\end{equation}  
The best two planet orbital parameters were employed as start coordinates for the combined HIRES/HARPS analysis. We also incorporated an additional unknown parameter $dc$ to allow for a possible difference in the constant velocity offsets of the HIRES and HARPS data. Based on our analysis of Figure~\ref{fig:velDiff}, our best estimate of $dc \approx 1.8$m s$^{-1}$. We assumed a Gaussian prior with zero mean and $\sigma = 3$m s$^{-1}$. 

\begin{figure}
\includegraphics[width=85mm]{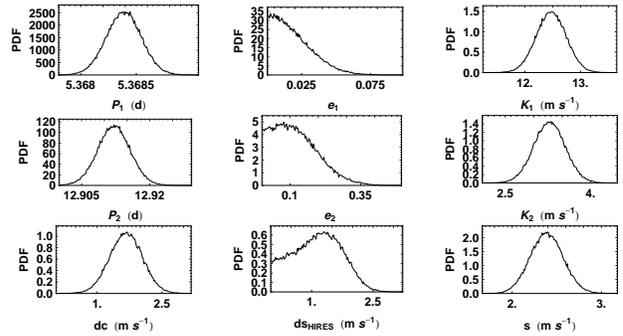}
\caption{A plot of a subset of the FMCMC parameter marginal distributions for a 2 planet fit of the combined HIRES/HARPS data.}
\label{fig:2planMargHIRESHARPS}
\end{figure}

The two planet FMCMC fit of the HIRES/HARPS data confirmed the 5.37 and 12.9d periods. Figure~\ref{fig:2planMargHIRESHARPS} shows a plot of a subset of the FMCMC parameter marginal distributions for the 2 planet fit of the data after filtering out the post burn-in FMCMC iterations that correspond to the 2 dominant period peaks at 5.37 and 12.9d. The bottom row shows the marginals for $dc$, $ds_{\rm{HARPS}}$, and $s$. The maximum a posterior and median values found for $dc$ are 1.65 and 1.67m s$^{-1}$, respectively, i.e., very close to the crude estimate of 1.8m s$^{-1}$ made earlier. 

\begin{figure}
\includegraphics[width=85mm]{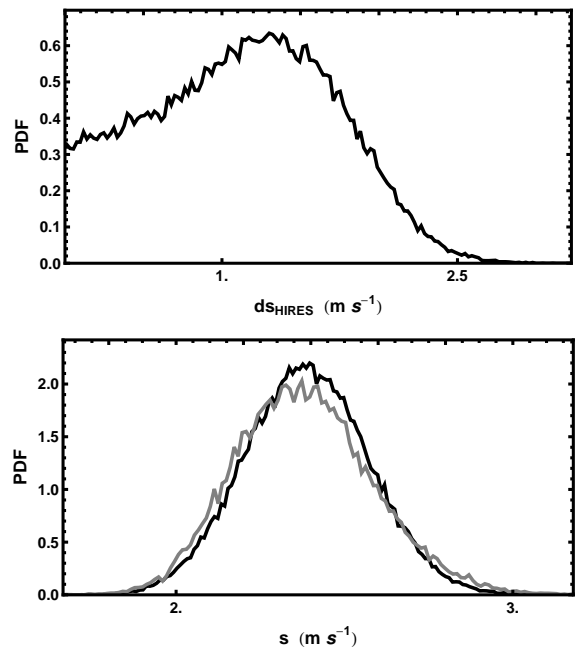}
\caption{The upper panel shows the marginal distribution for the unknown $ds_{\rm{HIRES}}$ parameter in the 2 planet FMCMC fit of the combined HIRES/HARPS data. The black curve in the lower panel shows the marginal distribution for the common extra noise parameter $s$ in the HIRES/HARPS fit. The light grey curve is the same quantity for the two planet HARPS only fit.}
\label{fig:2plansdsHIRESHARPS}
\end{figure}

The upper panel of Figure~\ref{fig:2plansdsHIRESHARPS} shows the marginal distribution for the unknown $ds_{\rm{HIRES}}$ parameter in the 2 planet FMCMC fit of the combined HIRES/HARPS data. The black curve in the lower panel shows the marginal distribution for the common extra noise parameter $s$ in the HIRES/HARPS fit. The light grey curve is the same quantity for the two planet HARPS only fit. Clearly there is very good agreement between the two $s$ parameter estimates.

\subsection{Three planet fit to the combined HIRES/HARPS data}
\label{sec:3HIRESHARPS}

In the above two planet fit to the combined HIRES/HARPS we found that the marginal distribution for the extra noise parameter $s$ agreed closely with that obtained from the two planet HARPS only data. For the three planet combined analysis we decided to fix the $s$ parameter to a value of 1.9m s$^{-1}$, i.e., the value obtained from the HARPS alone 3 planet fit. We decided to use the following noise model for the $j^{\rm{th}}$ data point for the three planet analysis.
\begin{equation}
\sigma_{\rm{HARPS}_j} = \sqrt{\sigma_{\rm{quoted}_j}^2 + 1.9^2}
\label{eq:sig3HARPS}
\end{equation} 
\begin{equation}
\sigma_{\rm{HIRES}_j} = \sqrt{\sigma_{\rm{quoted}_j}^2 + ds_{\rm{HIRES}}^2+ 1.9^2}
\label{eq:sig3HIRES}
\end{equation}  
Again the best three planet orbital parameters from the HARPS analysis were employed as start coordinates for the combined HIRES/HARPS analysis. All 3 period parameters were free to roam within a search range extending from 1.1d to $10 \times$ the data duration.  

\begin{figure}
\includegraphics[width=85mm]{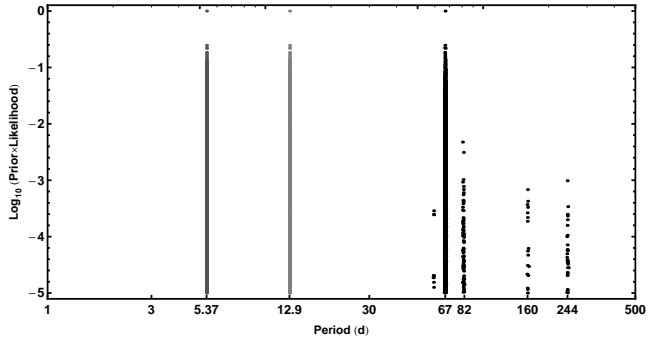}
\caption{A plot of the 3 period parameter values versus a normalized value of Log$_{10}$[Prior $\times$ Likelihood] for the 3 planet FMCMC Kepler fit of the combined HIRES/HARPS data.}
\label{fig:3planPerHIRESHARPS}
\end{figure}

Figure~\ref{fig:3planPerHIRESHARPS} shows plot of a sample of the FMCMC three planet periodogram. Only values within 5 decades of the maximum Log$_{10}$[Prior $\times$ Likelihood] are plotted but without regard to whether the values occurred before or after burn-in. Three prominent periods were clearly detected: 5.37, 12.9, and 66.9d. The third period parameter exhibited other peaks but these were significantly less probable, the one at 82d is consistent with being a one year alias of the 66.9d period. 

\begin{figure}
\includegraphics[width=85mm]{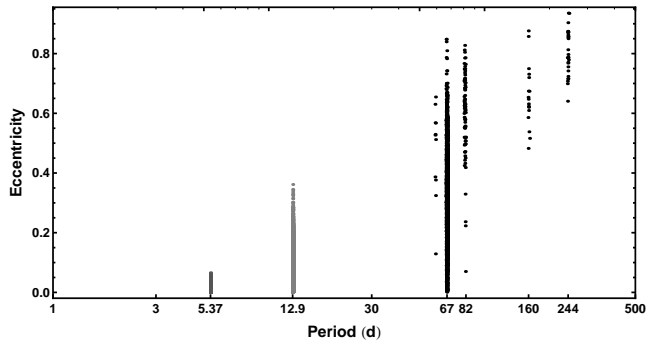}
\caption{A plot of eccentricity versus period for the 3 planet FMCMC fit of the combined HIRES/HARPS data.}
\label{fig:3planEccPHIRESHARPS}
\end{figure}

Figure~\ref{fig:3planEccPHIRESHARPS} shows a plot of eccentricity versus period for a sample of the FMCMC parameter samples for the 3 planet model. The dominant 5.37, 12.9, and 66.9d peaks allow for low eccentricity orbits. 

\begin{figure}
\includegraphics[width=85mm]{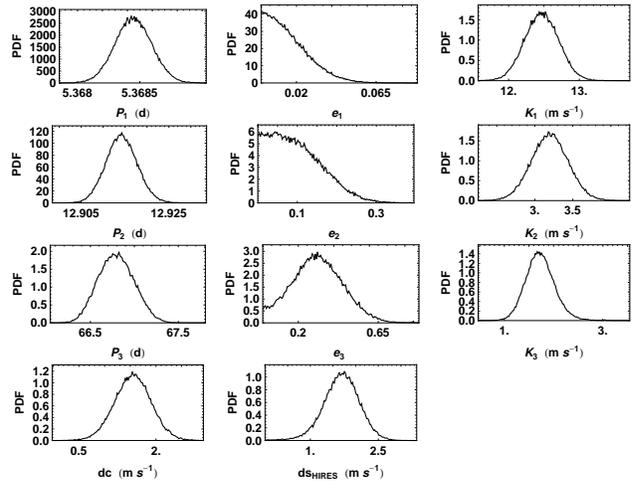}
\caption{A plot of a subset of the FMCMC parameter marginal distributions for a 3 planet fit of the combined HIRES/HARPS data.}
\label{fig:3planMargHIRESHARPS}
\end{figure}

Figure~\ref{fig:3planMargHIRESHARPS} shows a plot of a subset of the FMCMC parameter marginal distributions for the 3 planet fit of the data after filtering out the post burn-in FMCMC iterations that correspond to the 3 dominant period peaks at 5.37, 12.9, and 66.9d. The bottom row shows the marginals for $dc$, $ds_{\rm{HIRES}}$. The $ds_{\rm{HIRES}}$ is more accurately defined than in the two planet analysis. The maximum a posterior and median values found for $ds_{\rm{HIRES}}$ are 1.61 and 1.69m s$^{-1}$, respectively. 

\subsection{Four planet fit to the combined HIRES/HARPS data}
\label{sec:4HIRESHARPS}

For the four planet fit we reverted to the noise model used for the two planet analysis.
\begin{equation}
\sigma_{\rm{HARPS}_j} = \sqrt{\sigma_{\rm{quoted}_j}^2 + s^2}
\label{eq:sigHARPS}
\end{equation} 
\begin{equation}
\sigma_{\rm{HIRES}_j} = \sqrt{\sigma_{\rm{quoted}_j}^2 + ds_{\rm{HIRES}}^2+ s^2}
\label{eq:sigHIRES}
\end{equation}  
The best four planet orbital parameters from the HARPS only analysis were employed as start coordinates for the combined HIRES/HARPS analysis. As before we incorporated the additional unknown parameter $dc$ to allow for a possible difference in the constant velocity offsets of the HIRES and HARPS data. 

\begin{figure}
\includegraphics[width=85mm]{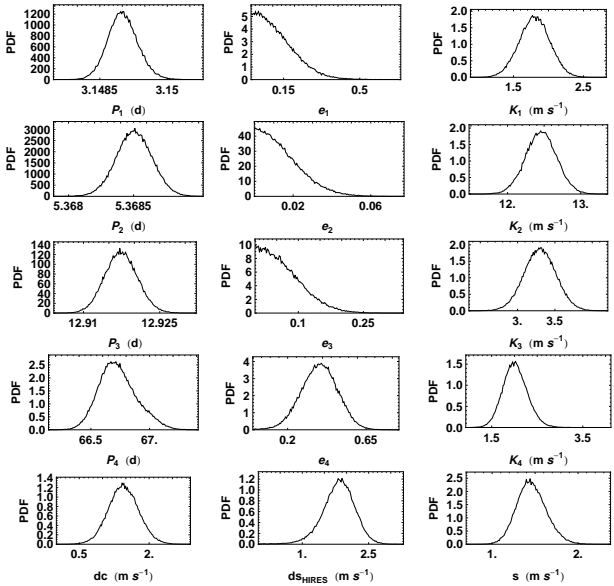}
\caption{A plot of a subset of the FMCMC parameter marginal distributions for a 4 planet fit of the combined HIRES/HARPS data.}
\label{fig:4planMargHIRESHARPS}
\end{figure}

The four planet Kepler periodogram found the four starting periods of 3.15, 5.37, 12.9, and 66.9d and no other peaks. Figure~\ref{fig:4planMargHIRESHARPS} shows the marginal posterior densities of a subset of the parameters.  

Table~\ref{tab:dcdss} shows a comparison of the estimates of the $dc$, $ds_{\rm{HIRES}}$, and $s$ parameters from the 2, 3, and 4 planet fits to the combined HIRES/HARPS data set. The parameter value listed is the median of the marginal probability distribution for the parameter in question and the error bars identify the boundaries of the 68.3\% credible region. The value immediately below in parenthesis is the MAP estimate, the value at the maximum of the joint posterior probability distribution.
\begin{table}
  \caption{Bayesian estimates of parameters $dc$, $ds_{\rm{HIRES}}$,
 and $s$ from the combined HIRES/HARPS data set for the 2, 3, and 4 planet fits.}
  \label{tab:dcdss}
  \begin{tabular}{@{}lllll@{}}
  \hline
   Parameter  & 2 planet & 3 planet & 4 planet \\
\hline
$dc$ & $1.67_{-0.35}^{+0.40}$ & $1.55_{-0.44}^{+0.25}$& $1.45_{-0.31}^{+0.32}$  \\
(m s$^{-1}$) & (1.64)& (1.65) & (1.53) \\
& & &  \\
$ds_{\rm{HIRES}}$ & $1.15_{-0.35}^{+1.05}$ & $1.69_{-0.32}^{+0.45}$ & $1.84_{-0.33}^{+0.35}$  \\
(m s$^{-1}$) & (0.75) & (1.61) & (1.85) \\
& & &  \\
$s$ & $2.39_{-0.11}^{+0.25}$ & $1.9$  & $1.34_{-.17}^{+.17}$   \\
(m s$^{-1}$) & (2.32) & fixed & (1.45) \\
& & &  \\
\hline
\end{tabular}
\end{table}
As expected the extra noise parameter $s$ decreases with the number of planets fit. The estimates of the offset parameter, $dc$, and the HIRES extra noise parameter, $ds_{\rm HIRES}$, agree within the quoted uncertainties and these uncertainties are smallest for the 4 planet model fit.  

We carried out an analysis of the 4 planet normalized fit residuals $(r_{1_j})$. We define $r_{1_j}$ in equ.~\ref{eq:4planReprResid}.
\begin{equation}
r_{1_j}=  \frac{1}{n_{\rm{i}}}\sum_{i=1}^{n_{\rm{i}}} \frac{{\rm residual}_{ij}}{\sqrt{\rm{error}_j^2 + ds_{{\rm{HIRES}}_i}^2 + s_i^2}}, 
\label{eq:4planReprResid}
\end{equation}
where $j$ is an index for the combined HIRES/HARPS data set. $n_i$ is the number of post burn-in FMCMC equilibrium samples used in computing the mean value of residual/(effective noise $\sigma$) for each measurement (typically we use $n_i \sim 200$). The effective noise $\sigma$ is given by
\begin{equation}
\rm{effective\ noise}\ \sigma= \sqrt{\rm{error}_j^2 + ds_{{\rm{HIRES}}_i}^2 + s_i^2}, 
\label{eq:4planNoiseSig}
\end{equation} 
which is the quadrature sum of the quoted error, $ds_{{\rm{HIRES}}_i}$, and the extra noise term $s_i$. Of course, $ds_{{\rm{HIRES}}_i}$ only contributes to the effective noise $\sigma$ of the HIRES data values. 
 
\begin{figure}
\includegraphics[width=85mm]{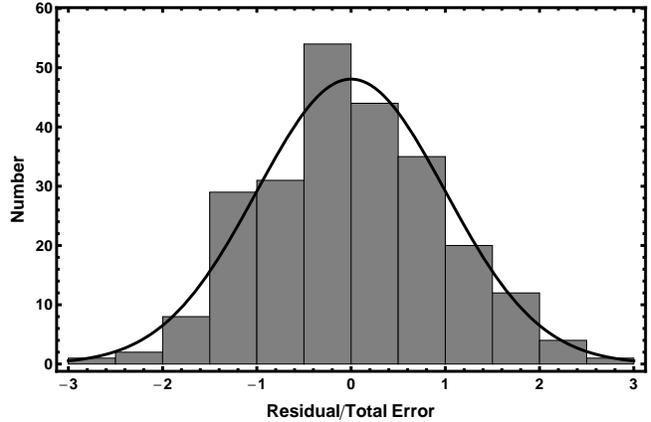}
\caption{A histogram of the $r_{1_j}$ values compared to a suitably normalized Gaussian with zero mean and standard deviation $=1$.}
\label{fig:4planReprResidHIRESHARPS}
\end{figure}

Figure~\ref{fig:4planReprResidHIRESHARPS} shows a histogram of the $r_{1_j}$ values compared to a suitably normalized Gaussian with zero mean and standard deviation $=1$. Within the uncertainties $r_{1_j}$ is consistent with a Gaussian distribution. 
The reduced $\chi^2$ of the $r_{1_j}$ is given by equ.~\ref{eq:4planChi2}.
\begin{equation}
\chi_{1}^2=  \frac{1}{n_t-n_p} \sum_{j=1}^{n_t} r_{1_j}^2=1.008, 
\label{eq:4planChi2}
\end{equation}
where $n_t =$ the total number of combined HIRES/HARPS data points and $n_p = 24$ is the number of fit parameters.

\begin{figure*}
\includegraphics[width=160mm]{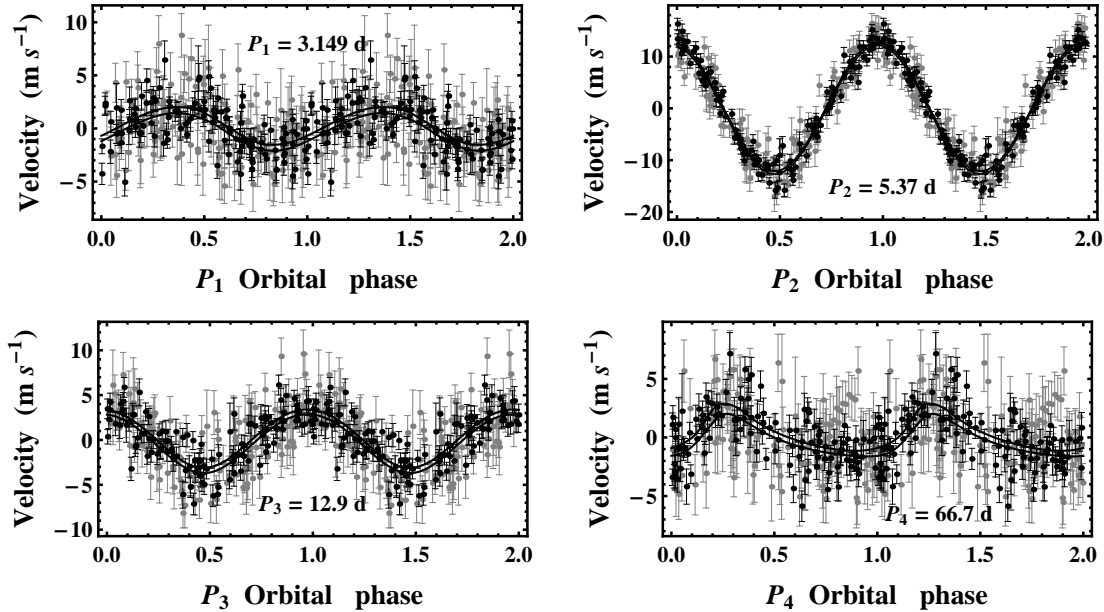}
\caption{Phase plots for the 4 most probable periods derived from the combined HIRES/HARPS data. The top left panel shows the data and model fit versus 3.15 day orbital phase after removing the effects of the five other orbital periods. The upper and lower curves are the mean FMCMC model fit $\pm 1$ standard deviation. The other three panels correspond to phase plot for the other three periods. The HARPS data points are black and the HIRES grey.}
\label{fig:4planPhaseHIRESHARPS}
\end{figure*}

\begin{figure}
\includegraphics[width=85mm]{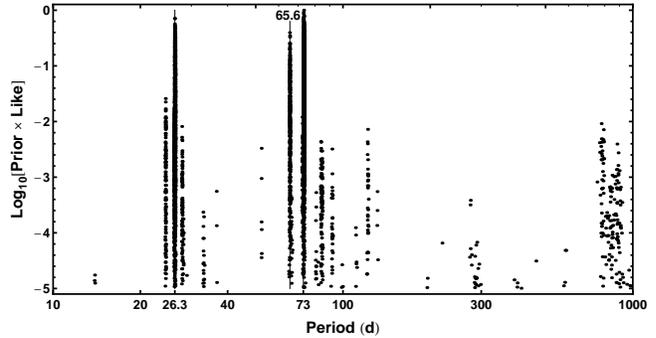}
\caption{A plot of the period parameter versus a normalized value of Log$_{10}$[Prior $\times$ Likelihood] for the 1 planet FMCMC Kepler fit to the HIRES residuals after removing the 3.15, 5.37, and 12.9 day orbits..}
\label{fig:1planPerHIRESresid}
\end{figure}

The four period phase plots are shown in Figure~\ref{fig:4planPhaseHIRESHARPS}. The top left panel shows the data and model fit versus 3.15d orbital phase after removing the effects of the three other orbital periods. The upper and lower solid curves are the mean FMCMC model fit $\pm 1$ standard deviation. The HIRES data points are shown in grey and the error bars are the quoted errors added in quadrature with our median estimate of $ds_{{\rm{HIRES}}_i}=1.84$m s$^{-1}$. The other panels correspond to phase plot for the other three periods.  
Examination of the 66.9d period phase plot indicates that the HIRES data do not lend much support to the 66.9 day period. In fact around a phase of 0.8 there are a series of $\sim 11$ points that exhibit a systematic trend away from the mean light curve. To examine this further we carried out a one planet FMCMC fit to the HIRES residuals after subtracting off the 3.15, 5.37, and 12.9d mean orbits. Figure~\ref{fig:1planPerHIRESresid} shows the one planet Kepler periodogram of these residuals. Three prominent peaks are present at 26.3, 65.6, and 73d along with many minor peaks. The thin solid vertical lines highlight these peaks. The strongest peak has a period of 73d which explains the absence of good support for a 66.9d period in Figure~\ref{fig:4planPhaseHIRESHARPS}.   

\subsection{Five planet fit to the combined HIRES/HARPS data}
\label{sec:5HIRESHARPS}

\begin{figure}
\includegraphics[width=85mm]{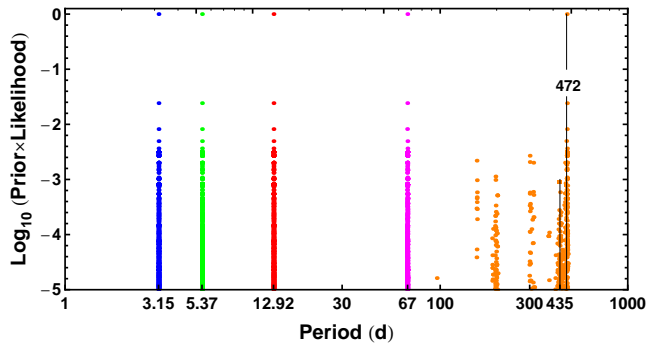}
\caption{A plot of the 5 period parameters versus a normalized value of Log$_{10}$[Prior $\times$ Likelihood] for the 5 planet FMCMC Kepler fit to the combined HIRE/SHARPS data.}
\label{fig:5planPerHIRESHARPS}
\end{figure}

\begin{figure}
\includegraphics[width=85mm]{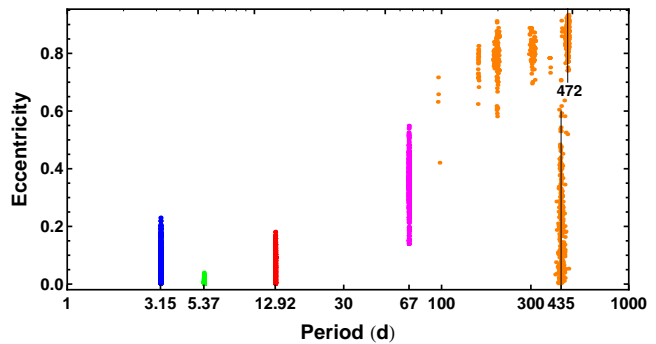}
\caption{A plot of eccentricity versus period for the 5 planet FMCMC fit to the combined HIRE/SHARPS data.}
\label{fig:5planEccPHIRESHARPS}
\end{figure}

Figures~\ref{fig:5planPerHIRESHARPS} and \ref{fig:5planEccPHIRESHARPS} show the results of a five planet fit to the combined HIRES/HARPS data set. The evidence for a $\sim 400$d period is more confused than the HARPS only results shown in Figures~\ref{fig:5planP}, \ref{fig:5planEccP}. The combined HIRES/HARPS results show a dominant high eccentricity peak with a period of 472d. Again, weak high eccentricity peaks are a common characteristic of noise. 

\section{Discussion}
\label{sec:discussion}

\begin{figure}
\includegraphics[width=85mm]{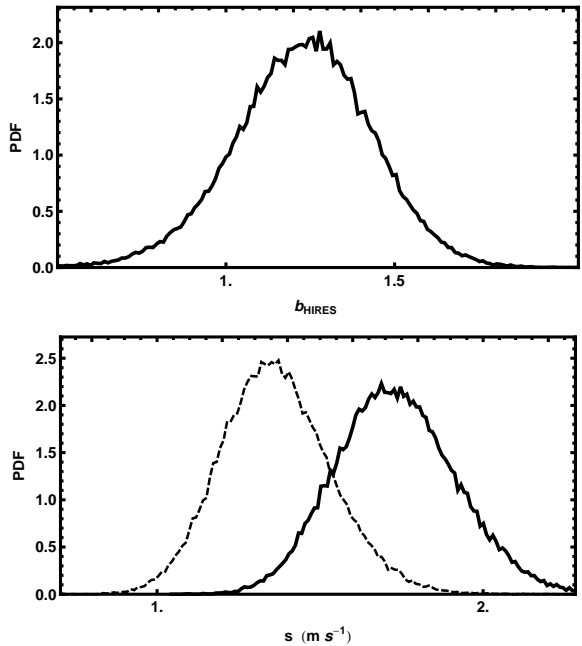}
\caption{A plot of the marginal probability distributions for the $b_{\rm{\rm HIRES}}$ and $s$ parameters for a 4 planet FMCMC fit to the combined HIRE/SHARPS data using the noise model of equ.~\ref{eq:bsigHIRES}. The dashed curve in the lower panel is the marginal for $s$ from the 4 planet fit to the HARPS only data.}
\label{fig:4planMargsb}
\end{figure}

Our analysis of the combined HIRES/HARPS data set argues strongly that the HIRES errors have been systematically underestimated. If we model this by the additive extra noise term $ds_{\rm{HIRES}}$, as in equ.~\ref{eq:bsigHIRES}, we conclude $ds_{\rm{HIRES}}= 1.84_{-0.33}^{+0.35}$. An alternative possibility that we can test is that the HIRES data are systematically too low by a common factor. To check this out we re-did the four planet analysis of the combined HIRES/HARPS data using the following noise model.
\begin{equation}
\sigma_{\rm{HIRES}_j} = \sqrt{(b_{\rm{HIRES}} \times \sigma_{\rm{quoted}_j})^2 + s^2}
\label{eq:bsigHIRES}
\end{equation}  
We assumed a uniform prior for the unknown parameter $b_{\rm{HIRES}}$ in the range 0.5 to 4.0.
The results were qualitatively very similar to the results obtained with $ds_{\rm{HIRES}}$ noise term. Following similar calculations to those outlined in Section~\ref{sec:4HIRESHARPS} we computed the reduced $\chi^2$ of the residuals divided by the total effective noise $\sigma$ and obtained a value of 1.009, indistinguishable from the reduced $\chi^2$ obtained using the $ds_{\rm{HIRES}}$ parameter. Figure~\ref{fig:4planMargsb} shows the marginal probability distributions for $b_{\rm{\rm HIRES}}$ and $s$ parameters. The dashed curve in the lower panel is the marginal for $s$ from the 4 planet fit to the HARPS only data. Employing the $b_{\rm{\rm HIRES}}$ parameterization results in a significantly larger estimate in the $s$ term than was required by the HARPS only analysis or the $ds_{\rm{\rm HIRES}}$ parameterization of the combined HIRES/HARPS data. 

Although the two different parameterizations of the HIRES extra noise lead to similar values of the reduced $\chi^2$ we can gain additional insight about which parameterization is better from a more microscopic exploration by binning the 4 planet fit residuals/(effective noise) and examining the $\chi^2$ values of the individual bins. If we add two $\chi^2$ random variables one of which has $
\nu_1$ degrees of freedom and the other has $\nu_2$ degrees of freedom then their sum will be $\chi^2$ with $\nu_1+\nu_2$ degrees of freedom, e.g., see \citealt{Gregorybook} p. 144. Similarly, we can take the quantity $r_{1_j}$ of equ.~\ref{eq:4planChi2} which is $\chi^2$ with $(n_t-n_p)$ degrees of freedom and divide the $r_{1_j}$ values into $k$ bins. Suppose the $k^{\rm{th}}$ bin has $n_k$ $r_{1_j}$ values then the $\chi^2$ of these values will be given by equ.~\ref{eq:kbinChi2}.
\begin{equation}
\chi_{k}^2=  \frac{1}{\frac{n_k}{n_t} \times (n_t-n_p)} \sum_{j=1}^{n_k} r_{1_j}^2, 
\label{eq:kbinChi2}
\end{equation}
which is $\chi^2$ with $\frac{n_k}{n_t} \times (n_t-n_p)$ degrees of freedom. 

\begin{figure}
\includegraphics[width=85mm]{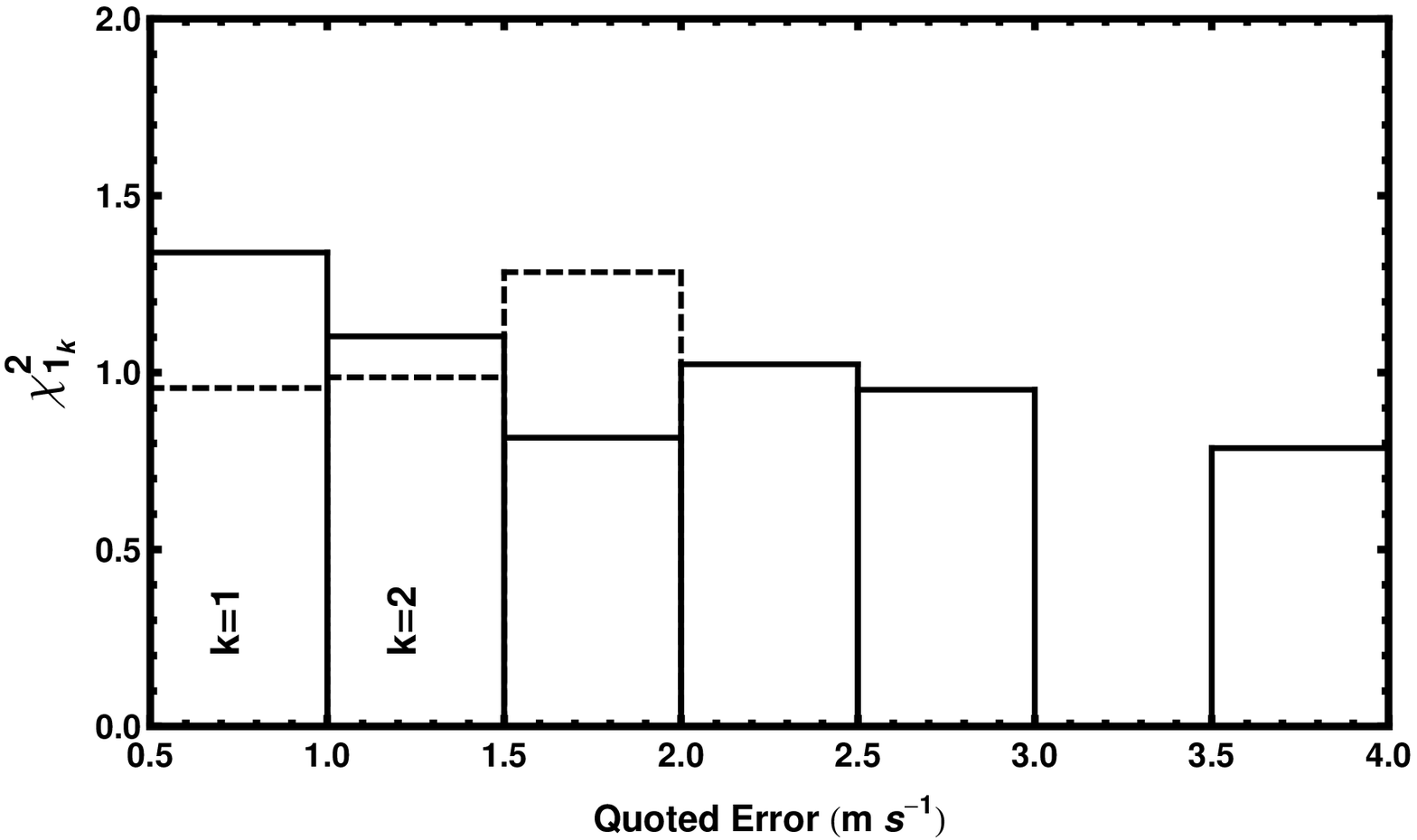}
\caption{A plot of the $\chi_k^2$ statistic from equ.~\ref{eq:kbinChi2} versus binned values of the quoted errors for the 4 planet fit of the combined HIRES/HARPS data using the $ds_{\rm{HIRES}}$ parameterization. The solid and dashed histograms show the binned HIRES data and binned HARPS data, respectively.}
\label{fig:4planChi2HIRESHARPS}
\end{figure}

\begin{figure}
\includegraphics[width=85mm]{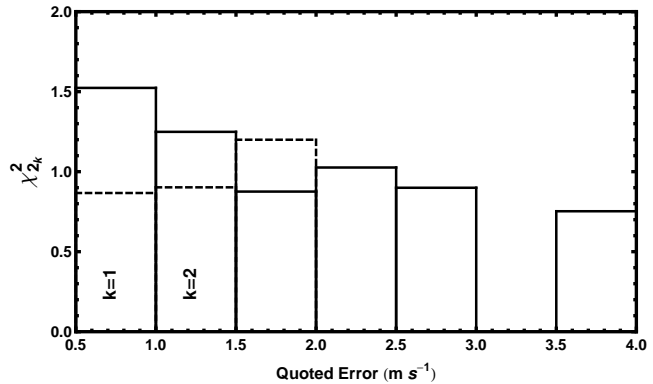}
\caption{A plot of the $\chi_k^2$ statistic from equ.~\ref{eq:kbinChi2} versus binned values of the quoted errors for the 4 planet fit of the combined HIRES/HARPS data using the $b_{\rm{HIRES}}$ parameterization. The solid and dashed histograms show the binned HIRES data and binned HARPS data, respectively.}
\label{fig:4planChi2bHIRESHARPS}
\end{figure}

Figure~\ref{fig:4planChi2HIRESHARPS} shows a plot of the $\chi^2$ statistic versus the original quoted errors that have been binned into 0.5m s$^{-1}$ bins for the 4 planet fit to the combined HIRES/HARPS data using the $ds_{\rm{\rm HIRES}}$ parameterization. Figure~\ref{fig:4planChi2bHIRESHARPS} shows a similar plot for the $b_{\rm{\rm HIRES}}$ parameterization. Ideally the $\chi^2$ distribution would be flat with a value of 1.0. It is clear that the $ds_{\rm HIRES}$ parameterization achieves a flatter distribution than the $b_{\rm{HIRES}}$ parameterization. Further the noise model involving $ds_{\rm{HIRES}}$ leads to a jitter noise estimate $s$ which is much closer to the value expected from the HARPS only 4 planet analysis.

Our HARPS only analysis provides evidence for 5 planetary signals while incorporating the HIRES data seems to degrade the evidence for both a 66.9 and $\sim 400$d periods even when we allow for an extra HIRES noise term of order 1.8m s$^{-1}$. 
We suspect this extra noise term may arise from an as yet unidentified systematics which may be the reason for the degradation in the quality of fits beyond three planets. Alternatively, could unidentified HARPS systematics be responsible for the extra periods evident in their lower noise data?

\section{Conclusions}

A Bayesian re-analysis of published HARPS and HIRES precision radial velocity data for Gl 581 was carried out with a multi-planet Kepler periodogram (from 1 to 6 planets) based on our fusion Markov chain Monte Carlo algorithm. In all cases the analysis included an unknown parameterized stellar jitter noise term. For the HARPS data set the most probable number of planetary signals detected is 5. The Bayesian false alarm probability for the 5 planet model is 0.01. These include the 3.15, 5.37, 12.9, 66.9d periods reported previously plus a $399_{-16}^{+14}$d period. The orbits of 4 of the 5 planets are consistent with low eccentricity orbits, the exception being the 66.9d orbit where $e=0.33_{-0.10}^{+0.09}$. The semi-major axis and $M sin i$ of the 5 planets are ($0.0285\pm0.0006$ au, $1.91_{-0.25}^{+0.26}$M$_{\earth}$), ($0.0406\pm0.0009$ au, $15.7_{-0.7}^{+0.7}$M$_{\earth}$), ($0.0730\pm0.0016$ au, $5.29\pm0.43$M$_{\earth}$), ($0.218\pm0.005$ au, $6.7\pm0.8$M$_{\earth}$), and ($0.72\pm0.24$ au, $6.6_{-2.7}^{+2.0}$M$_{\earth}$), respectively. 

In light of the \citet{Vogt2010} report of a sixth companion with a period of 36.6d, we carried out a 6 planet fit to the HARPS data which detected multiple period possibilities. The strongest of these, with a period $= 34.4\pm0.1$d and eccentricity of $0.49_{-0.17}^{+0.22}$, had a peak Log$_{10}$[Prior $\times$ Likelihood] 100 times larger than the others. The inferred semi-major axis and $M sin i$ are ($0.140\pm0.003$ au, $2.3_{-0.7}^{+0.8}$M$_{\earth}$). The Bayesian false alarm probability for the six planet model is extremely large $0.999978$ so we are unable to support any claim for a sixth companion on the basis of the current data.    
 
The analysis of the HIRES data set yielded a reliable detection of only the strongest 5.37 and 12.9 day periods. The analysis of the combined HIRES/HARPS data again only reliably detected the 5.37 and 12.9d periods. Detection of 4 planetary signals with periods of 3.15, 5.37, 12.9, and 66.9d was only achieved by including an additional unknown but parameterized Gaussian error term added in quadrature to the HIRES quoted errors. The marginal probability density of the sigma for this additional HIRES Gaussian noise term has a well defined peak at $1.84_{-0.33}^{+0.35}$m s$^{-1}$. Phase plots indicate that incorporating the HIRES data seems to degrade the evidence for the 66.9 and $\sim 400$d periods even when we allow for the extra HIRES noise term.  We suspect this extra noise term may arise from unidentified systematics which may be the reason for the degradation in the quality of fits beyond three planets. Alternatively, could unidentified HARPS systematics be responsible for the extra periods evident in their lower noise data? Independent experimental confirmation of the HARPS low noise level results would be very desirable.

\section*{Acknowledgments}

The author would like to thank Wolfram Research for providing a complementary license to run gridMathematica.

\bsp

\label{lastpage}

\end{document}